\documentclass[preprint, authoryear, 12pt]{elsarticle}

\usepackage{graphicx}
\usepackage{natbib}
\usepackage{amssymb}
\usepackage{subfigure}
\usepackage{color}
\usepackage{algorithm}
\usepackage{algpseudocode}
\usepackage{amsmath}

\def\bm#1{\mbox{\boldmath{$#1$}}}

\def \gdot{\bm{\nabla}\cdot}

\def \gxp{\xi\!\left(\left|\bm{x}-\bm{x}_p\right|\right)}
\def \epsf{\varepsilon_f}

\def \la{\langle}
\def \ra{\rangle}

\def \Vp{\mathcal{V}_p}

\DeclareMathAlphabet{\bbm}{U}{bbm}{m}{sl}

\journal{International Journal of Multiphase Flow}

\begin{document}

\begin{frontmatter}

\title{Transport modeling of sedimenting particles in a turbulent pipe flow using Euler-Lagrange large eddy simulation}

\author[cornell]{Sunil K. Arolla\corref{cor1}}
\ead{ska62@cornell.edu}

\author[cornell]{Olivier Desjardins}
\ead{olivier.desjardins@cornell.edu}

\address[cornell]{Sibley School of Mechanical and Aerospace Engineering, Cornell University, Ithaca, NY 14853, USA}

\cortext[cor1]{Corresponding author}

\begin{abstract}
A volume-filtered Euler-Lagrange large eddy simulation methodology is used to predict the physics of turbulent liquid-solid slurry flow through a horizontal pipe. A dynamic Smagorinsky model based on Lagrangian averaging is employed to account for the sub-filter scale effects in the liquid phase. A fully conservative immersed boundary method is used to account for the pipe geometry on a uniform cartesian grid. The liquid and solid phases are coupled through volume fraction and momentum exchange terms. Particle-particle and particle-wall collisions are modeled using a soft-sphere approach. A series of simulations are performed by varying the superficial liquid velocity to be consistent with the experimental data by \cite{dahl:2003}. Depending on the liquid flow rate, a particle bed can form and develop different patterns, which are discussed in light of regime diagrams proposed in the literature. The fluctuation in the height of the liquid-bed interface is characterized to understand the space and time evolution of these patterns. Statistics of engineering interest such as mean velocity, mean concentration, and mean streamwise pressure gradient driving the flow are extracted from the numerical simulations and presented. Sand hold-up calculated from the simulation results suggest that this computational strategy is capable of accurately predicting critical deposition velocity.
\end{abstract}

\begin{keyword}
liquid-solid slurry, sediment transport, bed formation, Euler-Lagrange large eddy simulation
\end{keyword}

\end{frontmatter}

\section{Introduction}\label{sec:intro}
Turbulent liquid-solid flows have a wide range of applications such as in the long distance transport of bulk materials to processing plants and in geomorphology where sediment may be entrained, transported, and deposited by water flow. Of particular interest in this work is the transport of oil sand through pipelines. In a near-horizontal pipeline, depending on the liquid flow rate, the slurry flow can exhibit four main regimes \citep{danielson:2007}. At low liquid flow rates, the sand sediments to the bottom of the pipe and forms a stable, stationary bed. When the liquid flow rate increases above a specific value, the sand starts getting transported in a thin layer above the bed. As the flow rate increases further, the sand bed breaks into a series of slow-moving dunes, which eventually grow to develop into a bed moving along the bottom of the pipe. At even higher flow rates, the sand particles ultimately become fully suspended in the carrier liquid. The velocity at the onset of a stationary bed formation is referred to as the critical deposition velocity. The formation of a stationary sand bed can pose several risks such as increased frictional losses, possibility of microbially-induced corrosion under the sand bed, and equipment failure due to sand accumulation. So, predicting the critical deposition velocity and understanding the physics of liquid-solid pipe flows is important for the efficient design of slurry pipelines. Moreover, the interaction between the turbulent carrier flow and a dense particle layer has great relevance in sediment transport modeling.

The response of a bed of particles to shearing flows has been widely studied in the context of the formation of alluvial river channels \citep{church:2006}. There are four dimensionless parameters that are used to parameterize the transport of sedimenting particles, the choice of which is non-unique \citep{yalin:1977}.  The first parameter that represents the incipient motion of particles is called the Shields number, $\theta={\tau_{b}}/{(\rho_{p}-\rho_{f})g d_p}$, where $\tau_{b}$ is the shear stress at the bed surface, $g$ is the acceleration due to gravity, $d_{p}$ is the particle diameter, and $\rho_{p}$ and $\rho_{f}$ are the density of the solid particles and fluid, respectively. This parameter measures the effect of the destabilizing hydrodynamic force over the stabilizing gravity. The threshold value for particle motion is denoted as the critical Shields number, $\theta_{c}$. The second parameter is the specific gravity of the solid particles, $s=\rho_{p}/\rho_{f}$. The third dimensionless parameter usually involves the fluid height, $h_{f}$, in the form of $h_{f}/d_{p}$, or a Froude number, $Fr=\sqrt{\tau_{b}/(\rho_{f}gh_{f})}$. The fourth parameter used is called the fall parameter, $R_{p}={d_{p}\sqrt{(s-1)gd_{p}}}/{\nu}$, which represents the relative importance of the gravity and $\nu$, the viscosity of the fluid \citep{jenkins:1998}. Experimental data collected in turbulent flows with sedimenting particles are conventionally represented in a \emph{Shields diagram} with Shields number as ordinate and fall parameter as abscissa. This curve is used to distinguish different modes of sediment transport as well as types of bedforms. When the flow is too weak to induce sediment motion ($\theta<\theta_{c}$), bedforms will be usually determined by previous stronger events \citep{nielsen:1992}. For flows such that $\theta_{c}<\theta<0.8$, bedforms such as vortex ripples or dunes will be present. For more intense flows with $\theta>0.8$, bedforms disappear and flat beds are observed in a regime called sheet flow. This criterion for sheet flow inception was proposed by \cite{wilson:1989}, but there are other criteria and formulas that exist in the literature. Most of the data available in the turbulent regime for threshold of particle motion present significant scatters due to systematic methodological biases of incipient motion of the bed \citep{buffington:1997,vanoni:1946,dancey:2002,paintal:1971}. Threshold values determined from bedload transport rate are usually larger than those deduced from visual observations of particle motion. This discrepancy is largely due to differences between the initial state of the bed as erosion and deposition are very sensitive to bed packing conditions \citep{papanicolaou:2002}. This is particularly important to acknowledge when validating computational methods against the data. Computer simulations should also exhibit such systematic bias provided the packed granular bed region is modeled. Presumably simpler laminar flows also suffer from the same difficulty. Recent work of \cite{ouriemi:2007} and \cite{peysson:2009} attempted to address this by providing robust and reproducible experimental measurements to infer critical Shields number.

The type of bedform significantly influences the flow characteristics such as resistance, mixing properties and importantly, characteristics such as thickness of the transport layer. So, it is highly desirable to be able to predict the nature of bedforms from an engineering point of view. It is now generally accepted that the mechanism that destabilizes a flat sediment bed is the phase lag between the perturbation in bed height and the bottom shear stress \citep{charru:2006}. Linear stability analysis is often applied to the problem in order to predict the most unstable wavelength, but this approach is not found to be satisfactory, with pattern wavelength prediction sometimes off by an order of magnitude \citep{raudkivi:1997,langlois:2007,coleman:2009,ouriemi:2009a}. Experimentally, a closed pipe configuration is ideal for such investigations as it offers a confined, well-controlled flow. A phase diagram showing different dune patterns observed with a bed composed of spherical particles in a pipe flow was presented by \cite{ouriemi:2009b}. They pointed out that the dune formation is controlled by the Reynolds number while the incipient motion of particles is controlled by the Shields number.

With increasing computational resources and advancements in numerical methods, computational fluid dynamics (CFD) provides a unique opportunity to understand the physics of liquid-solid pipe flows for a range of liquid flow rates. The Navier-Stokes equations are the governing equations for the carrier liquid phase, but it is not practical to resolve a wide range of scales in a turbulent flow. Moreover, resolving the flow around each particle becomes overly expensive. Recently, \cite{capecelatro:2013a} developed an Euler-Lagrange large eddy simulation (LES) framework in which the background carrier flow is solved using a volume-filtered LES methodology while each particle is tracked for its position and velocity in a Lagrangian approach. This methodology showed unique capability to predict particle bed formation and excellent agreement with the experimental data was obtained for a case with the liquid velocity above the critical deposition velocity \citep{capecelatro:2013b}. This framework can be used to explore the physics of liquid-solid slurry flows at Reynolds numbers of practical interest. To the best of our knowledge, no attempt to numerically simulate the evolution of bed of particles in a turbulent flow leading to pattern formation at high Reynolds numbers has been reported to date. Notably, \cite{kidanemariam:2014} presented results from fully resolved direct numerical simulations (DNS) in a channel flow configuration, but the highest Reynolds number based on bulk velocity is 6022. While their work demonstrated the feasibility of using fully resolved DNS for studying sediment pattern formation, the computational cost required was enormous and moreover, the computational domain size is too small to resolve the large scale features arising at high Reynolds numbers.

In this work, a volume-filtered Euler-Lagrange LES framework is applied to perform slurry flow simulations with different superficial liquid velocities and with the initial particle configuration set-up based on the sand hold-up data presented by \cite{danielson:2007},\cite{yang:2006}, and \cite{dahl:2003}. In contrast to the recent work by \cite{kidanemariam:2014}, the flow is resolved on a grid with cell size on the order of the particle diameter and the interphase momentum exchange is modeled via a drag term. However, a full description of contact mechanics is retained through the use of a soft-sphere collision model. In the previous work by \cite{capecelatro:2013b}, validation of Euler-Lagrange LES strategy above the critical deposition velocity is presented and the feasibility of using this method for predicting static bed formation is demonstrated. The specific objectives in this work are to further confirm the feasibility of this method in the context of turbulent transport of sedimenting particles, to compare predicted bedforms with those suggested by the data from the literature, and to validate global flow quantities such as mean streamwise pressure gradient and critical deposition velocity with the experimental data of \cite{dahl:2003}.


\section{Volume-filtered Euler-Lagrange LES framework} \label{sec:computation_framework}
To solve the equations of motion for the liquid phase without requiring to resolve the flow around individual particles, a volume filtering operator is applied to the Navier-Stokes equations, thereby replacing the point variables (fluid velocity, pressure, etc.) by smoother, locally filtered fields. The volume-filtered continuity equation is given by
\begin{equation}
\label{continuity}
\frac{\partial}{\partial t}\left(\varepsilon_{f} \rho_f\right)+\bm{\nabla} \cdot (\varepsilon_f \rho_f \bm{u}_f)=0,
\end{equation}
where $\epsf$, $\rho_f$, and $\bm{u}_f$ are the fluid-phase volume fraction, density, and velocity, respectively. The momentum equation is given by
\begin{equation}
\label{meanNS}
\frac{\partial}{\partial t}\left( \epsf \rho_f\bm{u}_f \right)+ \gdot \left(\epsf \rho_f \bm{u}_f \otimes \bm{u}_f\right)=
\gdot\left(\bm{\tau}-\bm{R}_{u}\right)+\epsf\rho_f \bm{g}-\bm{F}^\text{inter}+\bm{F}^\text{mfr},
\end{equation}
where $\bm{g}$ is the acceleration due to gravity, $\bm{F}^\text{inter}$ is the interphase exchange term that arises from filtering the divergence of the stress tensor, and $\bm{F}^\text{mfr} $ is a body force akin to a mean pressure gradient introduced to maintain a constant flow rate in the pipe. The volume-filtered stress tensor, $\bm{\tau}$, is expressed as
\begin{equation}
\label{stressbar}
\bm{\tau}=-p\bm{\mathcal{I}}+\mu\left[\nabla\bm{u}_f+\nabla\bm{u}_f^{\sf T}-\frac{2}{3}\ \left(\gdot\bm{u}_f\right)\bm{\mathcal{I}}\right]+\bm{R}_\mu,
\end{equation}
where the hydrodynamic pressure and dynamic viscosity are given by $p$ and $\mu$, respectively. $\bm{\mathcal{I}}$ is the identity tensor. $\bm{R}_\mu$ is an unclosed term that arises as a result of filtering the velocity gradients in the point wise stress tensor, and is modeled by introducing an effective viscosity $\mu^*$ to account for enhanced dissipation by the particles, given by
\begin{equation}
\bm{R}_{\mu} \approx \mu^*\left[\nabla\bm{u}_f+\nabla\bm{u}_f^{\sf T}-\frac{2}{3}\ \left(\gdot\bm{u}_f\right)\bm{\mathcal{I}}\right],
\end{equation}
where $\mu^*$ is taken from \cite{gibilaro:2007} for fluidized beds, and is given by
\begin{equation}
\label{effective_visc}
\mu^*=\mu\left(\varepsilon_{f}^{-2.8}-1\right).
\end{equation}
$\bm{R}_{u}$ is a sub-filter Reynolds stress term closed through a turbulent viscosity model, its anisotropic part is given by
\begin{equation}
\label{SGS}
\bm{R}_u-\frac{1}{3}tr(\bm{R}_{u}) \approx \mu_t\left[ \nabla\bm{u}_f+\nabla\bm{u}_f^{\sf T} \right],
\end{equation}
while the isotropic part is absorbed into pressure. A dynamic Smagorinsky model \citep{germano:1991,lilly:1992} based on Lagrangian averaging \citep{meneveau:1996} is employed to estimate the turbulent viscosity, $\mu_t$.

A Lagrangian particle-tracking approach is used for the solid phase. The displacement of an individual solid particle indicated by the subscript $p$ is calculated using Newton's second law of motion,
\begin{equation}
\label{newtonlaw}
m_p\frac{d\bm{u}_p}{dt}=\bm{f}^\text{inter}_p+\bm{F}^\text{col}_p+m_p\bm{g},
\end{equation}
where the particle mass is defined by $m_p=\pi\rho_p d_p^3/6$ and $\bm{u}_p$ is the particle velocity. The force $\bm{f}^\text{inter}_p$ exerted on a single particle $p$ by the surrounding fluid is related to the interphase exchange term in Eq.~\ref{meanNS} by
\begin{equation}
\label{Finterdef}
\bm{F}^\text{inter}=\sum_{p=1}^{n_p}\gxp\bm{f}^\text{inter}_p,
\end{equation}
where $n_p$ is the total number of particles, $\xi$ is the filtering kernel used to volume filter the Navier-Stokes equations, $\bm{x}_p$ is the position of the $p^{th}$ particle, and $\bm{f}^\text{inter}_p$ is approximated by
\begin{equation}
\label{fpinter2}
\bm{f}_p^\text{inter}\approx \Vp\gdot\bm{\tau}+\bm{f}_p^\text{drag},
\end{equation}
where $\Vp$ is the volume of the $p^{th}$ particle. The drag force is given as
\begin{equation}
\label{drag_acc}
\frac{\bm{f}_p^\text{drag}}{m_p}=\frac{1}{\tau_p}(\bm{u}_f-\bm{u}_p)F(\varepsilon_{f},{\rm Re}_p),
\end{equation}
where the particle response time $\tau_p$ derived from Stokes flow is
\begin{equation}
\label{response}
\tau_p=\frac{\rho_p d_p^2}{18\mu\varepsilon_f}.
\end{equation}
The dimensionless drag force coefficient $F$ of \cite{tenneti:2011} is employed in this work. Particle-particle and particle-wall collisions are modeled using a soft-sphere approach originally proposed by \cite{cundall:1979}. 

To simulate a fully-developed turbulent flow, periodic boundary conditions are used in the streamwise direction. In order to maintain a constant flow rate in this wall-bounded periodic environment, momentum is forced using a uniform source term that is adjusted dynamically in Eq.~\ref{meanNS}. 

To study the detailed mesoscale physics of slurries in horizontal pipes, the mathematical description presented heretofore is implemented in the framework of the NGA code \citep{desjardins:2008}. The Navier-Stokes equations are solved conservatively on a staggered grid with second order spatial accuracy for both the convective and viscous terms, and the second order accurate semi-implicit Crank-Nicolson scheme is employed for time advancement. The details on the mass, momentum, and energy conserving finite difference scheme are given by \cite{desjardins:2008}. The particles are distributed among the processors based on the underlying domain decomposition of the liquid phase. For each particle, its position, velocity, and angular velocity are solved using a second-order Runge-Kutta scheme. Coupling between the liquid phase and solid particles appears in the form of the volume fraction, $\epsf$, and interphase exchange term, $\bm{F}^\text{inter}$. These terms are first computed at the location of each particle, using information from the fluid, and are then transferred to the Eulerian mesh. To interpolate the fluid variables to the particle location, a second order trilinear interpolation scheme is used. To extrapolate the particle data back to the Eulerian mesh in a computationally efficient manner that is consistent with the mathematical formulation, a two-step mollification/diffusion operation is employed. This strategy has been shown to be both conservative and convergent under mesh refinement \citep{capecelatro:2013a}. A proper parallel implementation makes simulations consisting of $\mathcal{O}(10^8)$ Lagrangian particles and more possible, allowing for a detailed numerical investigation of slurries with realistic physical parameters. 

The liquid-phase transport equations are discretized on a uniform cartesian mesh, and a conservative immersed boundary (IB) method is employed to model the cylindrical pipe geometry without requiring a body-fitted mesh. The method is based on a cut-cell formulation that requires rescaling of the convective and viscous fluxes in these cells, and provides discrete conservation of mass and momentum. Details on coupling the IB method with the Lagrangian particle solver are given by \cite{capecelatro:2013a}.

\section{Flow configuration and simulation set-up}
\begin{figure}
\begin{minipage}[t!]{68mm}
\centering{
\subfigure[Sand hold-up data, from \cite{dahl:2003}.]{\includegraphics[width=60mm]{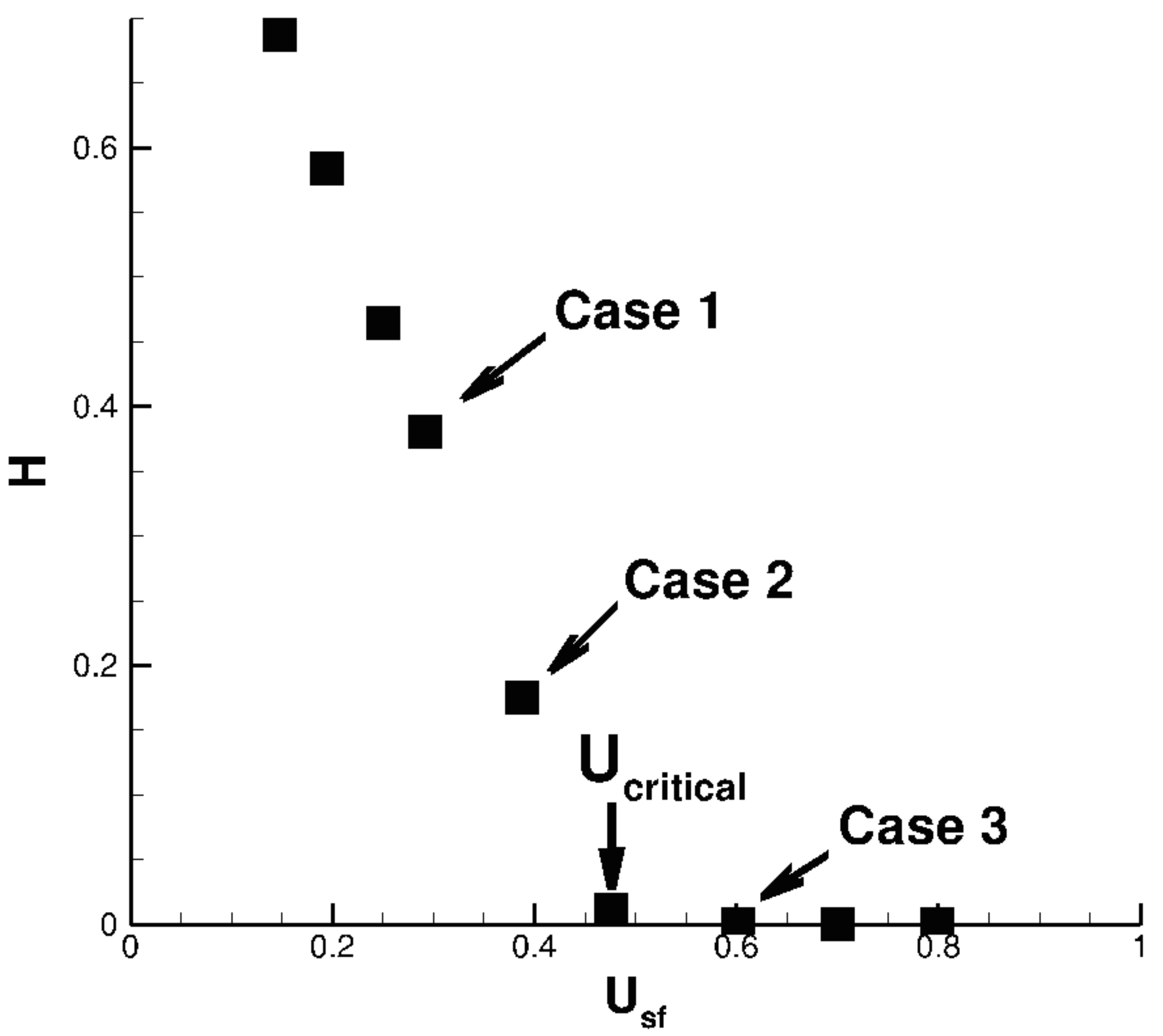} \label{cases_exp}}
}
\end{minipage}
\hfill
\begin{minipage}[t!]{80mm}
\centering{
\subfigure[Schematic showing different variables used to characterize the slurry flow.]{\includegraphics[width=70mm]{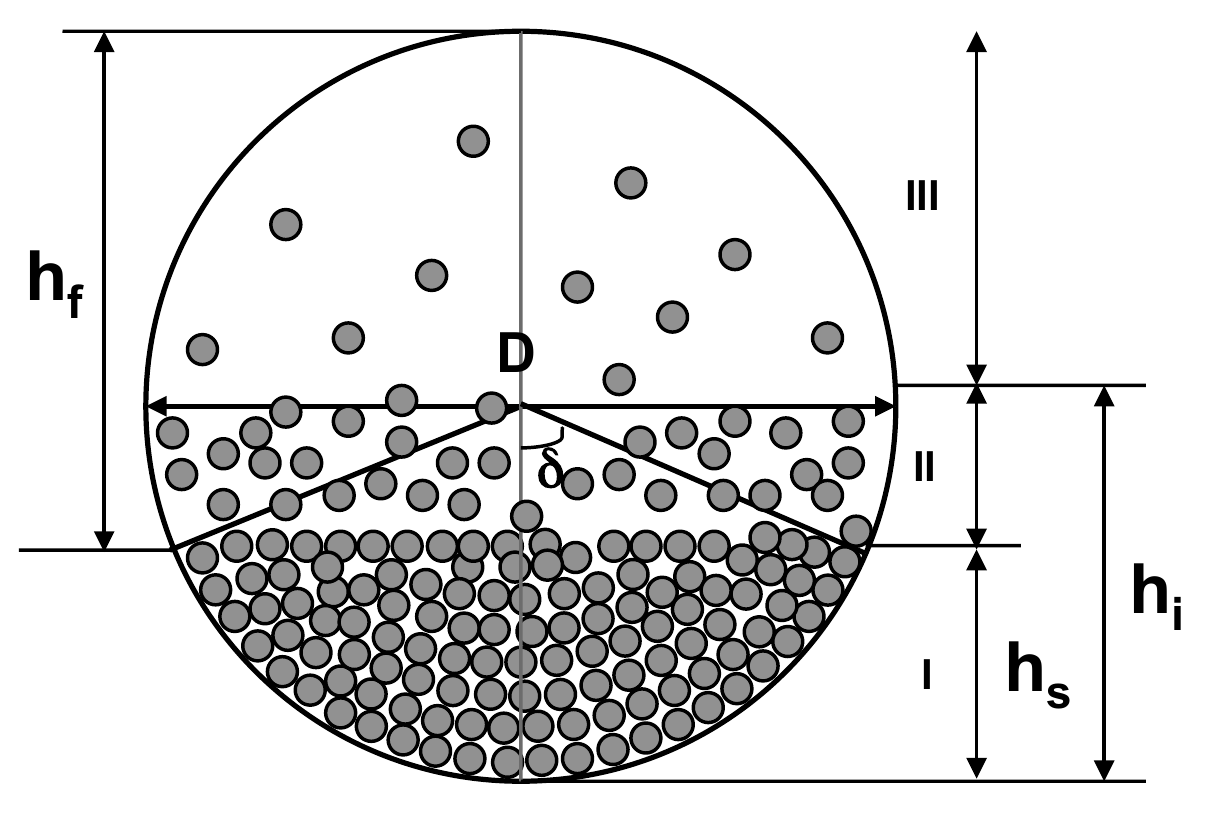} \label{holdup}}
}
\end{minipage}
\caption{Sand hold-up data from the experiments by \cite{dahl:2003}. The three cases considered for the simulations are shown by arrows. $U_{critical}$ is the critical deposition velocity and $h_{f}$ is the fluid height. Three distinct regions, as observed by \cite{capecelatro:2013b}, are identified: region I denotes the static bed, region II denotes an intermediate layer with erosion and deposition as dominant processes, and region III denotes the upper part of the pipe with few, fully entrained particles. Note that height of the static bed region is denoted as $h_{s}$ and the height of the liquid-bed interface is denoted as $h_{i}$. For exact definitions of these heights, see section \ref{globalStats}.}
\label{data_exp}
\end{figure}

The configuration considered in this work is that of a liquid-solid flow through a horizontal pipe. The experimental data reported in \cite{dahl:2003} form the basis for these simulations. In the experiments, sand is injected at a rate of 2.2 g/s. While the sand grain median diameter is provided, a full size distribution is not reported. Therefore, we assume that the sand particles are monodispersed with the diameter equal to the median diameter given in the experiments. The properties of the sand particles and of the fluid are given in table \ref{tab:parameters_a}. Available data include sand hold-up at different liquid flow rates as shown in figure \ref{data_exp} and the streamwise pressure gradient driving the flow. 

\begin{table}[!htp]
  \begin{center}
\def~{\hphantom{0}}
  \begin{tabular}{c c}
\hline
       Pipe diameter, $D$                     & 0.069 m \\
       Particle diameter, $d_{p}$              & $280\times10^{-6}$ m \\
       Particle density, $\rho_{p}$            & 2650 kg/m$^{3}$ \\
       Liquid density,   $\rho_{f}$            & 998 kg/m$^{3}$ \\
       Particle-particle coefficient of restitution      & 0.9 \\
       Particle-wall coefficient of restitution          & 0.8 \\
       Coefficient of friction                           & 0.1 \\
\hline
  \end{tabular}
  \caption{Flow parameters}
  \label{tab:parameters_a}
  \end{center}
\end{table}

To set-up the numerical simulations consistently with these experiments, the number of particles are computed using the sand hold-up data as follows. The sand hold-up is defined as percentage area of the pipe cross-section occupied by the static bed. It is calculated as
\begin{equation}\label{holdup_eqn}
H=\frac{1}{2\pi}\left[2\delta-\sin(2\delta) \right],
\end{equation}
where $\delta$ is the angle defined as $\delta=\cos^{-1}(1-2h_{s}/D)$, where $D$ is the pipe diameter given in table \ref{tab:parameters_a} and $h_{s}$ is the static bed height. Both variables are used to characterize the static bed region, referred to as region I in figure \ref{holdup}. The volume fraction of the particles averaged over the entire volume of the pipe is given by 
\begin{equation}
\la \varepsilon_{p} \ra_{pipe}=0.63 H +\la \varepsilon_{p} \ra_{ab} (1-H),
\end{equation}
where $\la \varepsilon_{p} \ra_{ab}$ denotes the mean volume fraction above the bed. Note that the static bed is assumed to be at the random close packing limit. It is further assumed that the particles above the bed are transported at the bulk velocity of the liquid. This is a crude assumption, but below the critical deposition velocity ($U_{critical}$) relatively fewer particles get resuspended, so the error induced will be small. This, however, leads to larger errors when the liquid velocity is above $U_{critical}$, where most particles are resuspended. With this approximation, $\la \varepsilon_{p} \ra_{ab}$ can be computed as
\begin{equation}
\la \varepsilon_{p} \ra_{ab} \approx \frac{\dot{m}_{p}}{\rho_{p}U_{sf}A},
\end{equation}
where $\dot{m}_{p}$ is the experimental sand injection rate of 2.2 g/s, $A$ is the cross-sectional area of the pipe, and $U_{sf}$ is the superficial liquid velocity. The number of particles can then be calculated from
\begin{equation}
\la \varepsilon_{p} \ra_{pipe} \frac{\pi}{4}D^{2}L_{x} = N_{p} \frac{\pi}{6} d_{p}^{3},
\end{equation}
where the $L_x$ is the domain size in the streamwise direction.

The number of particles calculated above are randomly distributed in the computational domain. To obtain initial conditions, the simulation is first run by disregarding the hydrodynamic forces. This allows only the gravity and particle-particle collision forces to act on the particles, which eventually results in a static bed. Then the fluid-particle interaction is activated by specifying a superficial liquid velocity in accordance with the experiments. As the liquid flow rate increases, the particles get eroded from the static bed and get transported along with the carrier liquid. Using the mean particle velocity, the particle mass flow rate can be computed and compared with the sand injection rate as a  consistency check.

\begin{table}[!htp]
  \begin{center}
\def~{\hphantom{0}}
  \begin{tabular}{c c c c}
\hline
       Case & $U_{sf}$(m/s) & $Re_{sf}$ & $N_{p}$ \\[3pt]
\hline
         1  &      0.3      &   20700   &    $19.25 \times 10^{6}$  \\
         2  &      0.4      &   27600   &    $8 \times 10^{6}$  \\
         3  &      0.6      &   41400   &    $0.042 \times 10^{6}$\\
\hline
  \end{tabular}
  \caption{Different cases considered, see also figure \ref{cases_exp}. $N_{p}$ is the number of particles, $U_{sf}$ is the superficial liquid velocity in m/s, and $Re_{sf}$ is Reynolds number based on superficial velocity and pipe diameter.}
  \label{cases}
  \end{center}
\end{table}

\begin{figure}
\centering{
\includegraphics[width=90mm]{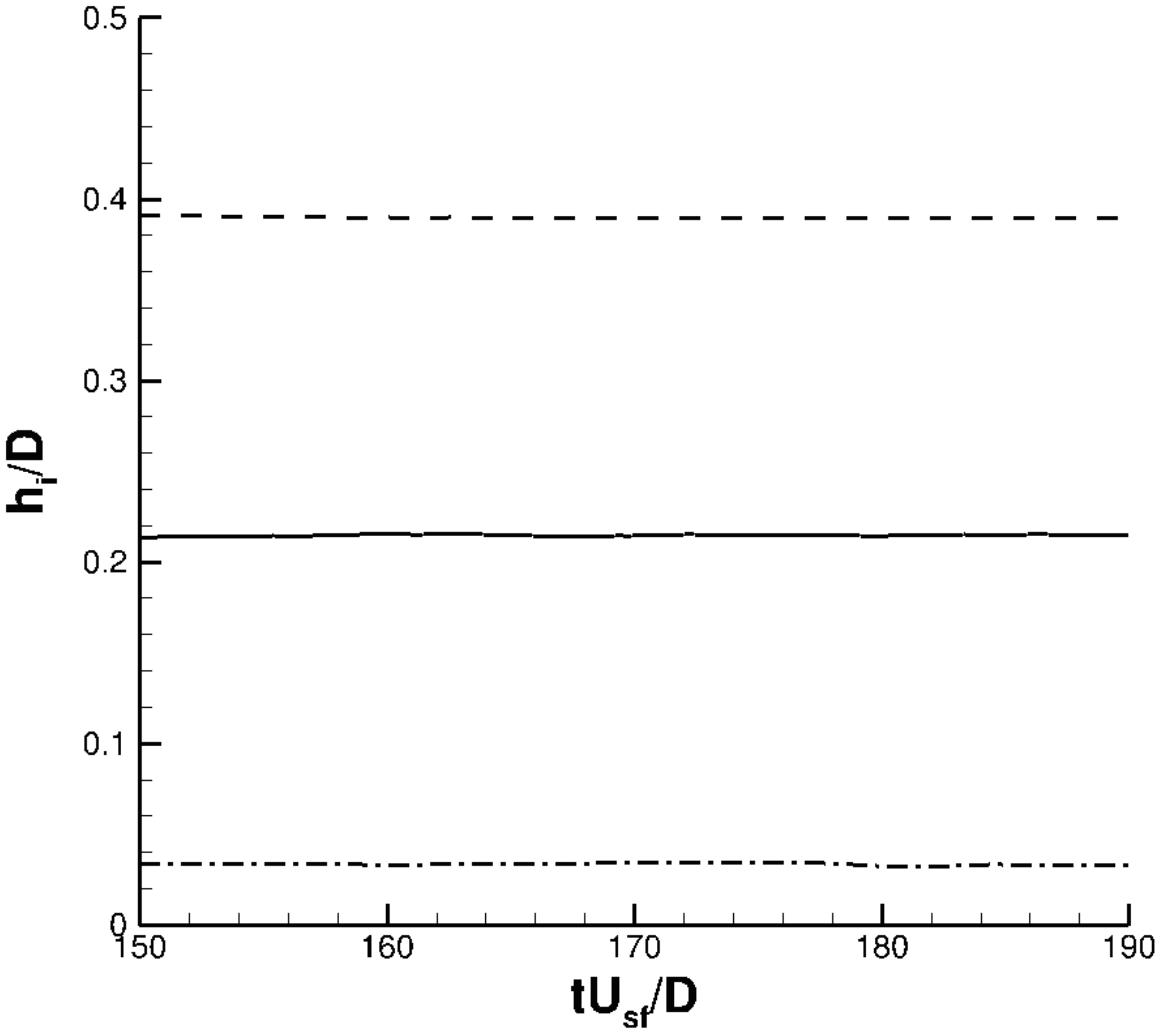}
}
\caption{Convergence history showing mean height of the liquid-bed interface as a function of non-dimensional time.}
\label{converge_hist}
\end{figure}

Due to the high computational cost of the simulations, our investigation is currently restricted to only three cases that assess the capability of Euler-Lagrange LES approach to predict the slurry flow physics. The particular cases considered are presented in table~\ref{cases}. The computational grid used has $768 \times 156 \times 156$ points in the streamwise ($x$), vertical ($y$), and lateral ($z$) directions, respectively. To capture the mesoscale features in this flow, each cell size is chosen to be equal to $1.57$ times the particle diameter. The domain length is chosen to be $L_x=3.75D$, resulting from a compromise between the need to resolve long streamwise flow structures and the need for the simulations to be computationally tractable. The simulations are run for approximately 150 inertial time scales given by $tU_{sf}/D$. The statistics are extracted over another 40 time scales. The convergence of the simulations in terms of the mean height of the liquid-bed interface, $h_{i}$, is shown in figure \ref{converge_hist}, which will be defined in section \ref{globalStats}.

\section{Results and discussion}
Due to the high cost of the simulations, the discussion of results will be focused on first-order statistics such as mean velocity and mean concentration.

\subsection{Global statistics}\label{globalStats}
First, the pressure gradient driving the flow from numerical simulations is compared with the experimental data from \cite{dahl:2003} in table \ref{tab:dpdx}. The agreement with the experiments is within approximately $10 \%$. This slight discrepancy can be attributed in part to the assumptions used in the configuration set-up. Table \ref{tab:dpdx} is also provides the average particle mass flow rate calculated from the simulations, which is found to be on the same order of magnitude, although systematically larger than the sand injection rate given in the experiments (2.2 g/s). The pressure gradient needed to drive the flow first decreases with decrease in $U_{sf}$. As $U_{sf}$ is decreased, the sand particles tend to deposit more at the bottom, which in turn ultimately leads to an increase in pressure gradient. This non-monotonic behavior of pressure gradient is well reproduced by the present simulation approach.

\begin{figure}
\centering{
\includegraphics[width=90mm]{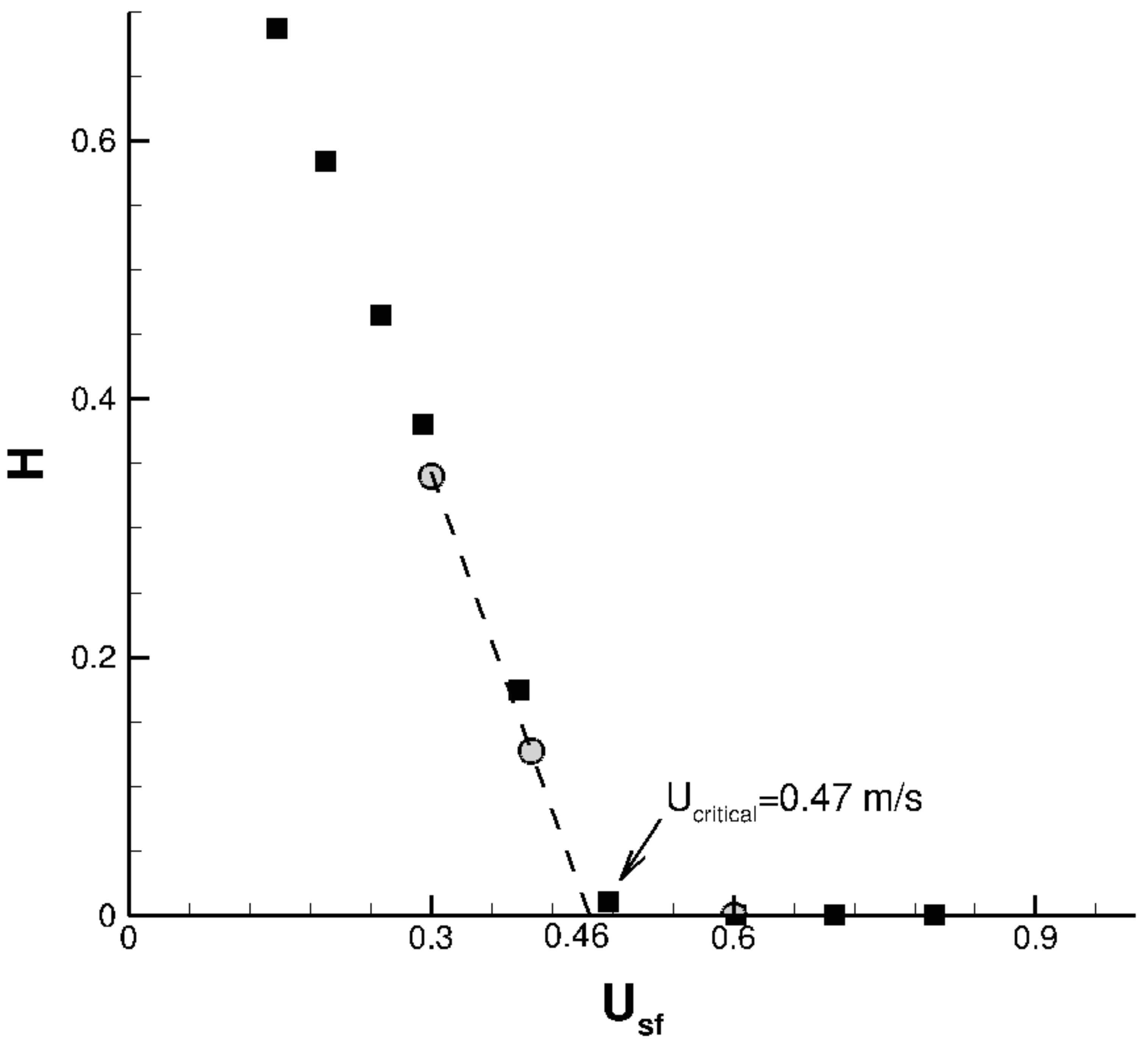}
}
\caption{Comparison of the sand hold-up predictions with the data. Squares: Experiments of \cite{dahl:2003}, circles: present simulations.}
\label{holdup_compare}
\end{figure}

\begin{table}[!htp]
  \begin{center}
\def~{\hphantom{0}}
  \begin{tabular}{c c c c c}
\hline
       Case & $U_{sf}$  & $(dp/dx)_{exp}$  & $(dp/dx)_{LES}$ & $\dot{m}_{p}$ \\[3pt]
\hline
         1  &      0.3       &  80.807         &   71.5          & 3.8                    \\
         2  &      0.4       &  61.741         &   70.4          & 3.2                    \\
         3  &      0.6       &  72.59          &   74.64         & 4.2                    \\
\hline
  \end{tabular}
  \caption{Pressure gradient at different superficial liquid velocities in Pa/m. $\dot{m}_{p}$ is the average particle mass flow rate in g/s.}
  \label{tab:dpdx}
  \end{center}
\end{table}
To further validate the three simulations, the sand hold-up is computed \emph{a posteriori} from the simulations using equation \ref{holdup_eqn}. The results are presented in table \ref{tab:holdup}, where it can be observed that the predicted values are in good agreement with the experimental data. Since we set up the simulations using the hold-up data and with crude assumptions on suspended particles, the hold-up is lower than the experimental value once the particles get resuspended. We predict zero hold-up for case 3 in agreement with the data. To predict $U_{critical}$ from the simulations, the data points below $U_{critical}$ are extrapolated linearly to the axis representing zero hold-up, as illustrated in figure \ref{holdup_compare}. This gives $U_{critical}$ from the simulations to be $0.46$ m/s, which is within $3\%$ of the experimental value.

\begin{table}[!htp]
  \begin{center}
\def~{\hphantom{0}}
  \begin{tabular}{c c c c c c}
\hline
       Case & $U_{sf}$(m/s) & $H_{exp}$ & $H_{LES}$  & $h_{s}$  & $h_{i}$ \\[3pt]
\hline
         1  &      0.3      &   0.362   &  0.341  & $92 d_{p}$  & $96 d_{p}$  \\
         2  &      0.4      &   0.150   &  0.127  & $45 d_{p}$  & $53 d_{p}$  \\
         3  &      0.6      &   0.000   &  0.000  & 0.0        &  $8 d_{p}$  \\
\hline
  \end{tabular}
  \caption{Sand hold-up comparison with experimental data.}
  \label{tab:holdup}
  \end{center}
\end{table}

Table~\ref{tab:holdup} also reports two characteristic heights that define the interface between the liquid and the particle bed. First, the height of the static bed is defined as
\begin{equation}
h_{s}(x,t)=\langle y \, |\, u_p = 0.01U_{sf}\rangle_z,
\end{equation}
where $\la \cdot \ra_{z}$ represents lateral averaging over five cells on both sides of the centerline of the pipe. A mean value for the static bed height $h_s$ can be obtained by performing averaging in the streamwise direction and in time, i.e., $h_s=\langle h_s\left(x,t\right)\rangle_{x,t}$. This corresponds to the height of region I shown in figure \ref{holdup} where all the particles are stationary and contact mechanics plays dominant role. As expected, the static bed height is found to decrease monotonically with superficial liquid velocity.

Following \cite{kidanemariam:2014}, a second definition for the height of the interface is calculated using a threshold value of particle volume fraction of 10\%, leading to
\begin{equation}
h_{i}(x,t)=\langle y \, |\, \varepsilon_p = 0.1\rangle_z.
\end{equation}
The mean liquid-bed interface height $h_i$ can be calculated by further averaging in the streamwise direction and in time, i.e., $h_i=\langle h_i\left(x,t\right)\rangle_{x,t}$. This corresponds to the height of the regions I and II combined as shown in figure \ref{holdup}. While particles are essentially static in region I, particles in region II are transported in a thin collisional layer where erosion and deposition are the dominant processes. This interface height is also found to decrease monotonically with superficial liquid velocity, but it remains larger than $h_s$ in all three cases considered. The thickness of region II, characterized by the difference between $h_{i}$ and $h_{s}$, is found to be $4d_{p}$ for case 1 and $8d_{p}$ for cases 2 and 3, which confirms the existence of a very thin transport layer at the surface of the static bed (or at the pipe surface for case 3 where a static bed is absent).



\subsection{Mean flow profiles}\label{mean}
The vertical profiles of the first-order statistics presented in this section are extracted at the pipe centerline ($z=0$), averaged in the streamwise direction and in time. 

\begin{figure}
\centering{
\includegraphics[width=90mm]{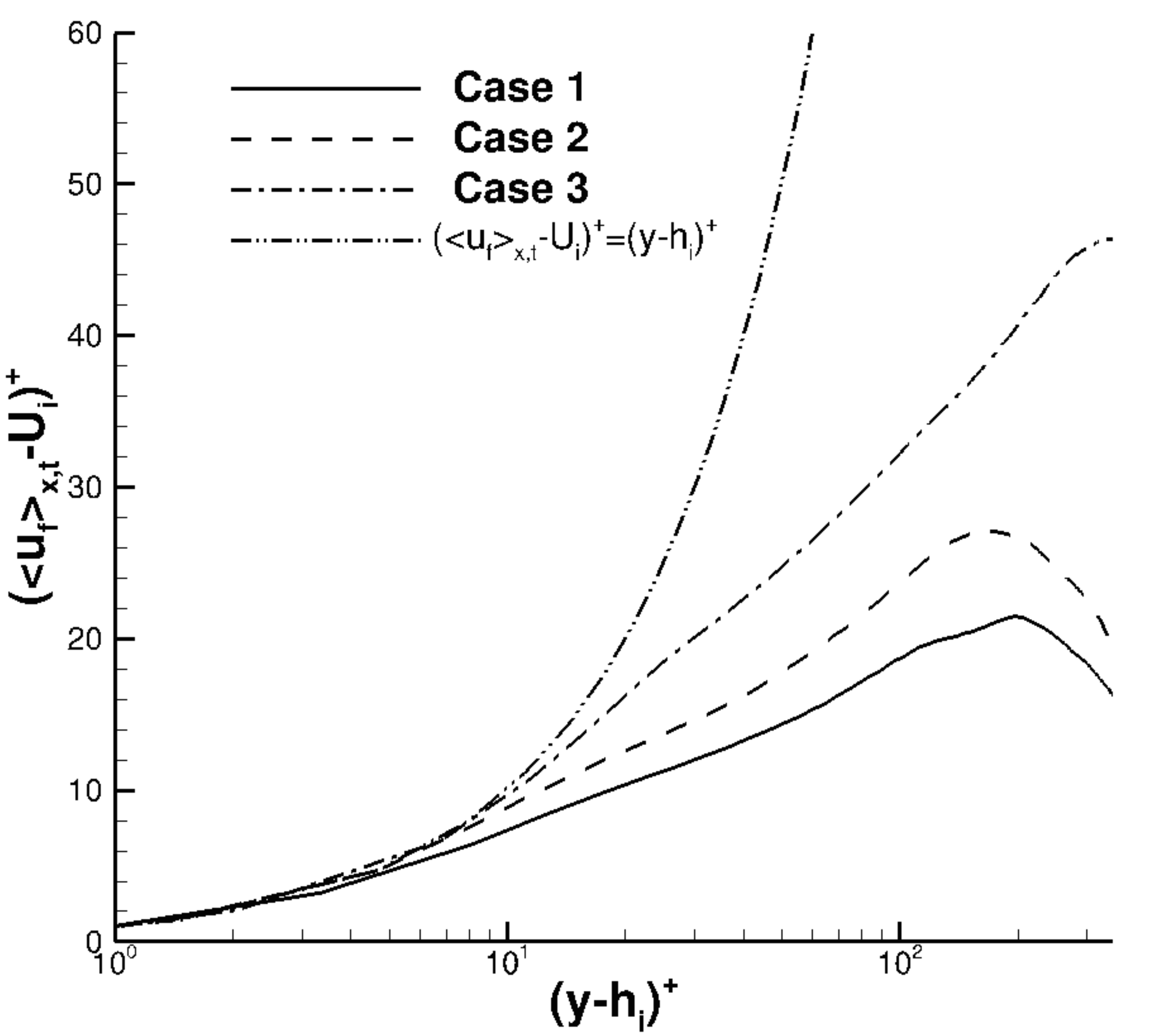}
}
\caption{Mean liquid velocity profiles plotted in interface plus units. $U_{i}$ is the mean liquid velocity at $h_{i}$.}
\label{meanLiquidVelocity}
\end{figure}

In annular liquid-gas flows, \cite{mccaslin:2014} showed that velocity profiles within the gas core preserve some characteristics of the law of the wall for single-phase pipe flow, when normalized appropriately. This is an interesting observation for modeling considerations. Motivated by this, the mean liquid velocity normalized in the units of liquid-bed interface are plotted in figure \ref{meanLiquidVelocity} from our current simulations. Note that the profiles are plotted in shifted coordinates using the liquid-bed interface height $h_{i}$ and the velocity is normalized using the friction velocity calculated at this interface. Due to the presence of dense particle layer above the bed, the apparent viscosity increases and turbulence is attenuated, leading to a distinguishable viscous sublayer for all the cases. However, a clear region where a logarithmic law holds is difficult to identify, although it seems to be present in case 3. Moreover, there is no self-similar behavior as the liquid phase Reynolds number is varied. This can potentially be explained by the fact that the confinement of the liquid flow by the particle bed varies strongly with superficial velocity, therefore the flow configuration itself is very different in all three cases.

\begin{figure}
\centering{
\includegraphics[width=90mm]{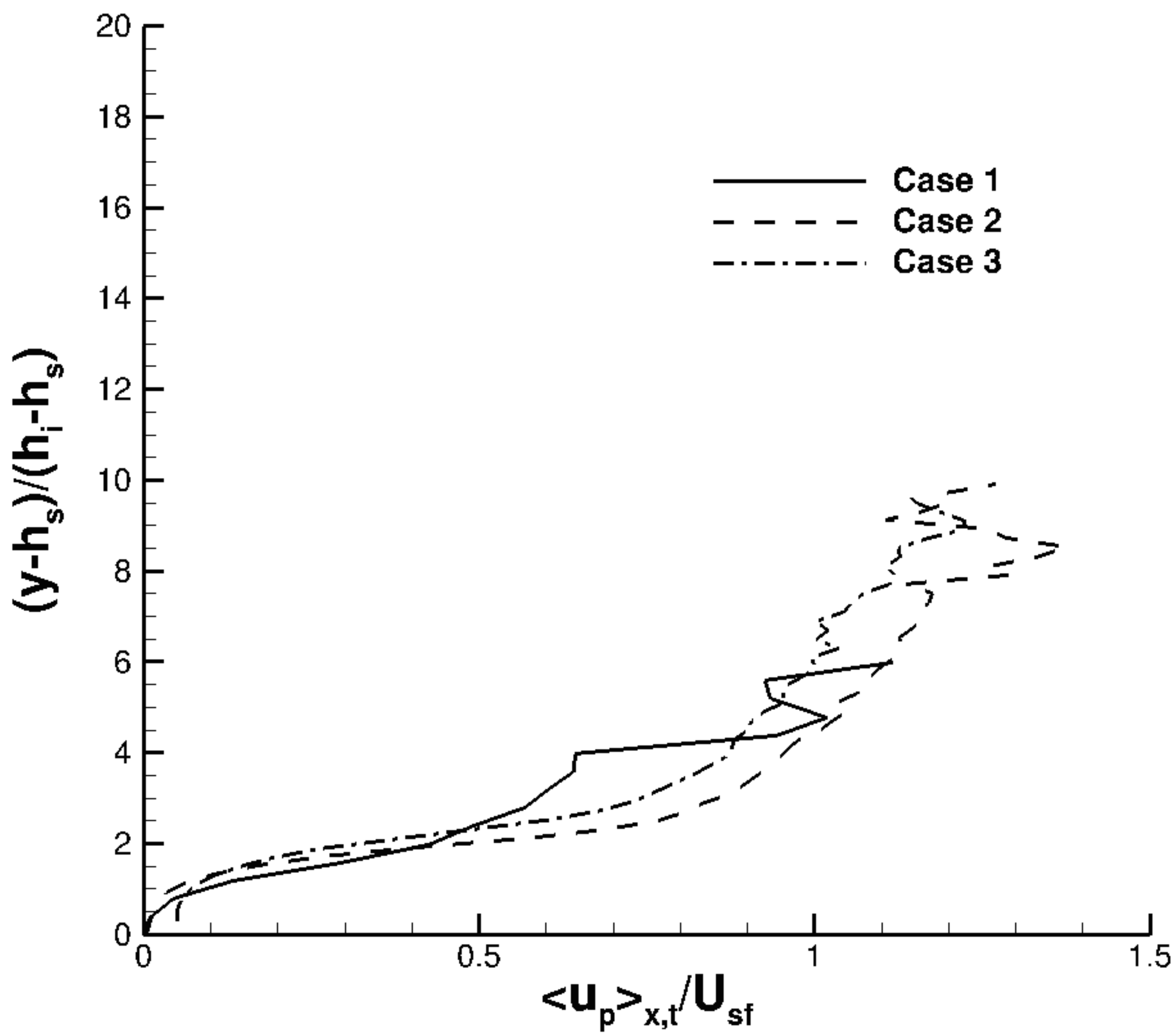}
}
\caption{Mean normalized particle velocity profiles as a function of normalized position within region II.}
\label{meanVelocity}
\end{figure}

\begin{figure}
\centering{
\includegraphics[width=90mm]{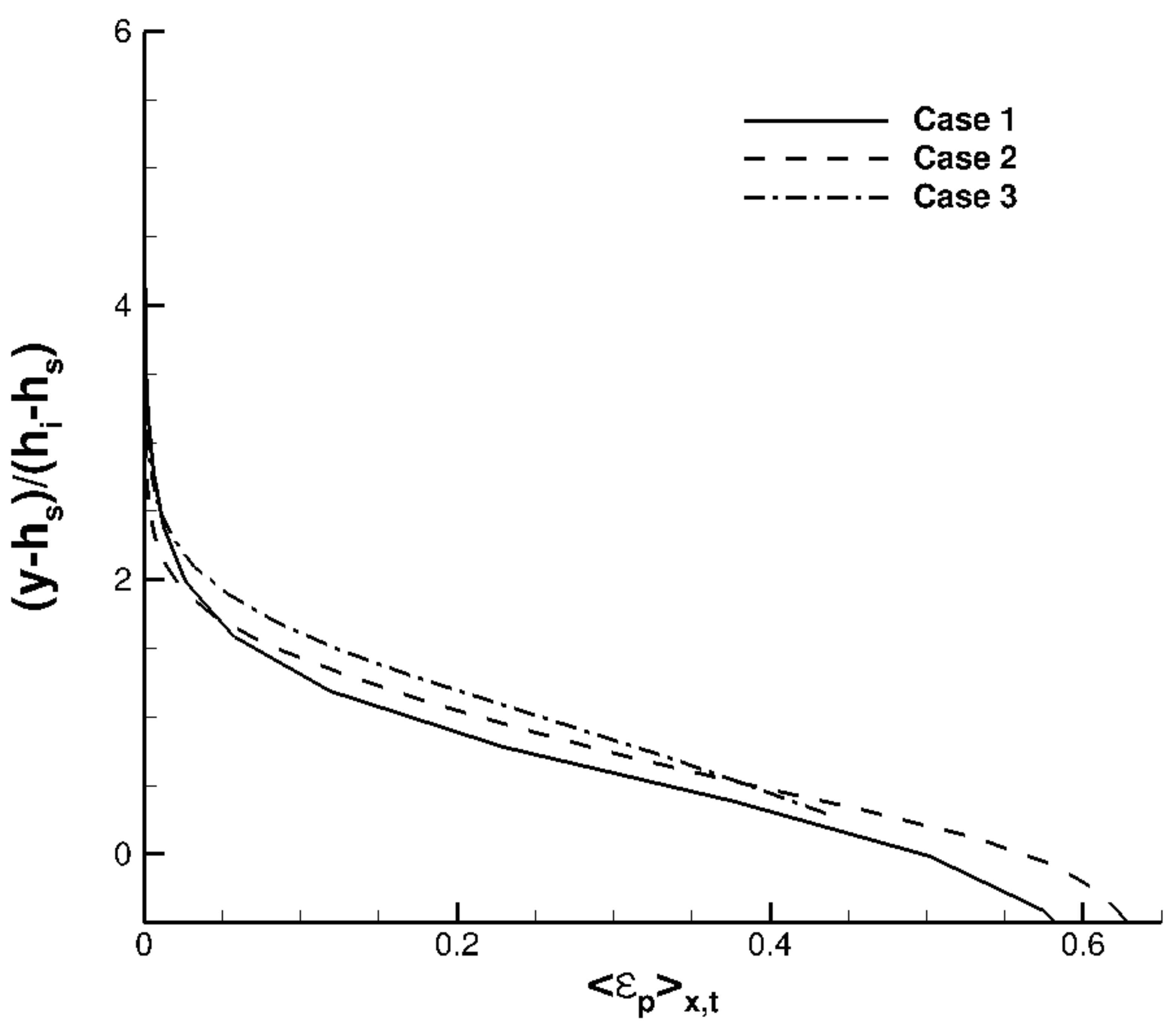}
}
\caption{Mean particle concentration profiles as a function of normalized position within region II.}
\label{meanConcentration}
\end{figure}

The mean particle velocity and concentration profiles shown in figures \ref{meanVelocity} and \ref{meanConcentration} are plotted in terms of $(y-h_s)/(h_i-h_s)$, the normalized position within region II. All concentration profiles show a good collapse, highlighting that in all cases, most moving particles are contained in region II. The velocity profiles also show a good collapse, although these profile extend much further than region II. This confirms that in region III (above region II), few particles are fully suspended with velocities on the order $U_{sf}$.

\subsection{Characterizing the transport layer thickness}
The thickness of the layer in which the particles are being transported is an important parameter from an engineering analysis point of view. Following \cite{duran:2012}, the characteristic transport layer thickness $\lambda$ can be defined as 
\begin{equation}
\lambda = \left(\frac{\int_{0}^{D}(y-\overline{y})^{2} \langle u_{p} \rangle_{x,t}(y)dy}{q} \right)^{1/2},
\end{equation}
where $\overline{y}=\frac{1}{q}\int_{0}^{D}y \langle u_{p} \rangle_{x,t} (y) dy$ gives the height of the transport layer center, $q=\int_{0}^{D} \langle u_{p} \rangle_{x,t} (y) dy$ is the volume flux of the particles, and $\la \cdot \ra_{x,t}$ denote averaging in the streamwise direction and in time. The transport layer thickness calculated is presented in table \ref{transportLayer}. This quantity confirms that transport happens in a thin layer on the order of few particle diameters, and that this transport layer is thinnest for the lowest liquid flow rate. Note that it varies non-monotonically with $U_{sf}$, probably because case 3 does not have a static bed. This transport layer thickness is larger in magnitude than the thickness representing region II given by $h_{i}-h_{s}$, which is also included in the table. This can be attributed to the fact that $\lambda$ measures transport deeper within the bed, instead of using a low velocity cut-off criterion. Despite their differences, both $\lambda$ and $h_{i}-h_{s}$ provide a consistent measure of transport layer thickness, and both show a similar trend when $U_{sf}$ is varied.

\begin{table}[!htp]
  \begin{center}
\def~{\hphantom{0}}
  \begin{tabular}{c c c c}
\hline
       Case & $U_{sf}$(m/s) & $\lambda$ & $h_{i}-h_{s}$\\[3pt]
\hline
         1  &      0.3      &   9$d_{p}$  & 4$d_{p}$  \\
         2  &      0.4      &   14$d_{p}$ & 8$d_{p}$  \\
         3  &      0.6      &   11$d_{p}$ & 8$d_{p}$  \\
\hline
  \end{tabular}
  \caption{Characterization of the transport layer thickness.}
  \label{transportLayer}
  \end{center}
\end{table}

\subsection{Pattern formation and phase diagrams}
\begin{figure}
\begin{minipage}[t!]{68mm}
\centering{
\subfigure[Shields diagram from sediment transport literature]{\includegraphics[width=68mm]{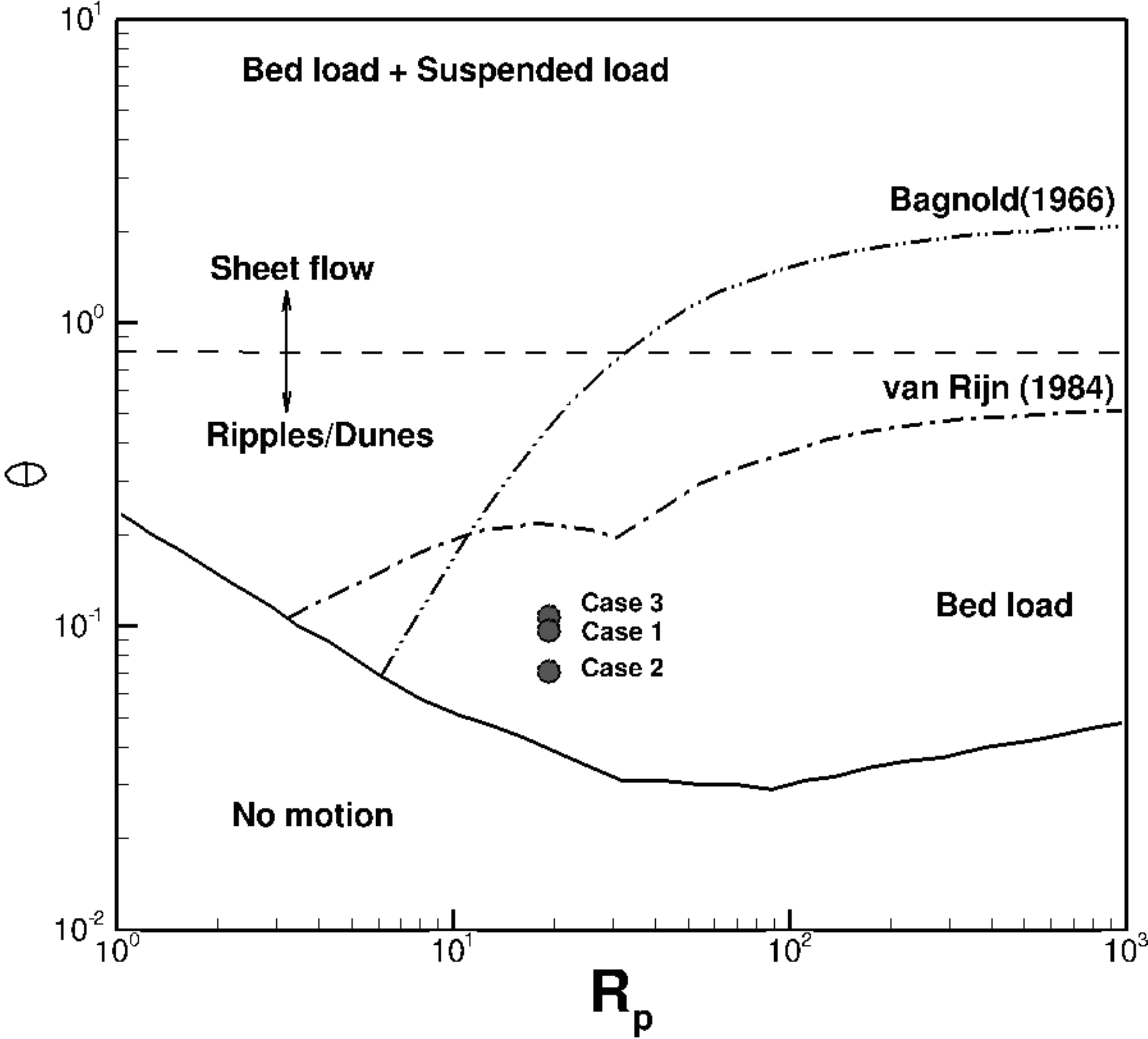}}
}
\end{minipage}
\hfill
\begin{minipage}[t!]{68mm}
\centering{
\subfigure[Phase diagram in $Re$-$Ga(h_{f}/d_{p})^{2}$ plane]{\includegraphics[width=68mm]{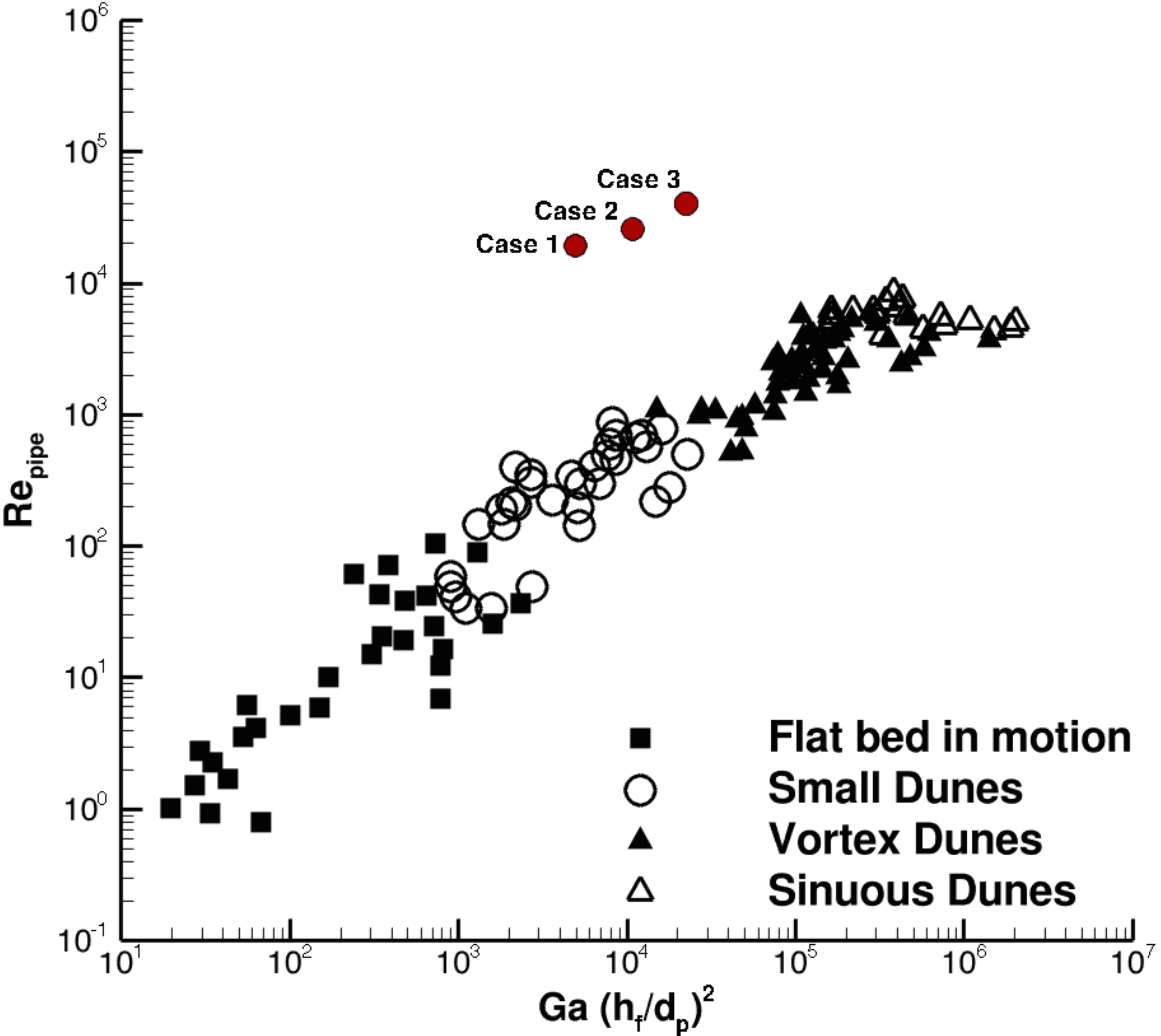}}
}
\end{minipage}
\caption{Phase diagrams from experimental observations.}
\label{phaseDiagrams}
\end{figure}

Another interesting question to address is the capability of phase diagrams presented in the literature to predict the pattern formation above the static bed. The formation of ripples/dunes is widely reported in the sediment transport literature in the form of Shields diagram. To represent the current simulations within this diagram, the Shields parameter $\theta$ needs to be calculated. The classical definition of $\theta$ is
\begin{equation}\label{shields_def1}
\theta = \frac{\tau_{b}}{(\rho_p-\rho_f)g d_p},
\end{equation}
where $\tau_{b}$ is the bed shear stress. From Euler-Lagrange simulations, $\tau_b$ can be inferred using the liquid velocity field along with the location of the liquid-solid interface, $h_i$, using
\begin{equation}
\tau_{b}=\mu^{*} \frac{\partial \la u_{f} \ra_{x,t}}{\partial y}\bigg{|}_{h_{i}},
\end{equation}
where $\la u_{f} \ra_{x,t}$ is the mean liquid velocity as shown in figure \ref{meanLiquidVelocity}. Note that the Shields number aims at providing an estimate of the relative magnitude of the streamwise and vertical forces, which is readily available in the present Lagrangian treatment of the particles. Therefore, the Shields number can be extracted directly from Lagrangian particle force data using
\begin{eqnarray}\label{shields_def2}
\theta=\langle f_{p,x}^\text{inter}\rangle_t/f_{p,y},
\end{eqnarray}
where $f_{p,y}$ denotes vertical force which is equivalent to the apparent weight of the particles. The streamwise force is extracted in a thin layer at the surface of the static bed. The Shields number computed using the Eulerian data using equation \ref{shields_def1} is denoted as $\theta_{e}$, while the Shields number computed using the Lagrangian data with equation \ref{shields_def2} is denoted as $\theta_{l}$. The calculated values for both parameters are presented in the table \ref{shields}, showing that $\theta$ can be calculated consistently using both expressions. Since the Shields number computed using the Lagrangian data directly represents the forces felt by the particles near the surface of the bed, we use this definition to place our simulations within the Shields diagram that is shown in figure \ref{phaseDiagrams}.

In figure~\ref{phaseDiagrams}, the Shields curve represented as a solid line denotes the threshold for incipient sediment motion. Curves showing \cite{bagnold:1966} and \cite{vanrijn:1984} denote initiation of the suspended load. Particular bedform is determined by the dotted line plotted at $\theta=0.8$, below which the data indicates formation of ripples/dunes. Note that all three simulations fall in this regime, hence dune formation is expected from the Shields diagram.

\begin{table}[!htp]
  \begin{center}
\def~{\hphantom{0}}
  \begin{tabular}{c c c c}
\hline
       Case & $U_{sf}$ (m/s) & $\theta_{e}$  &  $\theta_{l}$ \\ [3pt]
\hline
         1  &      0.3      & 0.0961       &   0.110       \\
         2  &      0.4      & 0.0722       &   0.071       \\
         3  &      0.6      & 0.1071       &   0.119       \\
\hline
  \end{tabular}
  \caption{Shields parameter calculated from Eulerian and Lagrangian data.}
  \label{shields}
  \end{center}
\end{table}

The Shields parameter is primarily meant for predicting the incipient motion of the sediment particles. Instead, \cite{ouriemi:2009b} plotted a phase diagram in the $Re$--$Ga(h_{f}/d_{p})^{2}$ plane to explain pattern formation specific to pipe flows. Here, the Galileo number is defined as $Ga=d_p^{3}(\rho_p-\rho_f)g/\nu$, and can be thought of as a Reynolds number based on the particle settling velocity, and is therefore analogous to the fall parameter $R_p$ on the Shields diagram. Note that these experiments are performed up to Reynolds number of 5,800, whereas the present simulations are conducted at significantly higher Reynolds numbers. With reference to the abscissa, the lowest liquid flow rate corresponds to small dunes regime, while at higher flow rates, ``vortex dunes'' are expected based on this diagram.

\begin{figure}
\begin{minipage}[t!]{70mm}
\centering{
\subfigure{\includegraphics[width=120mm]{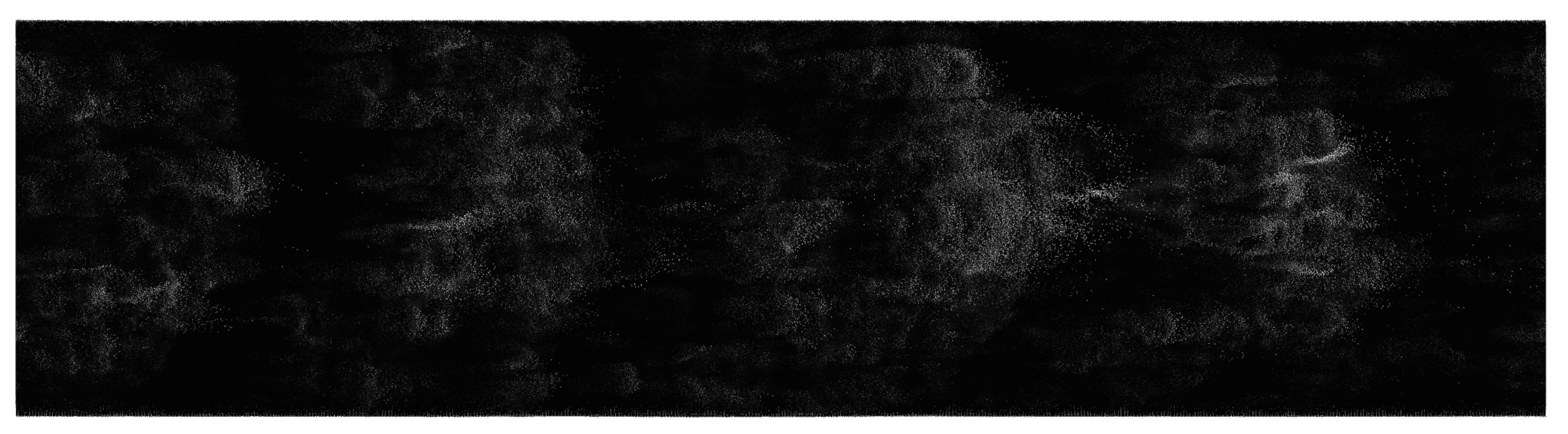}}
\subfigure{\includegraphics[width=120mm]{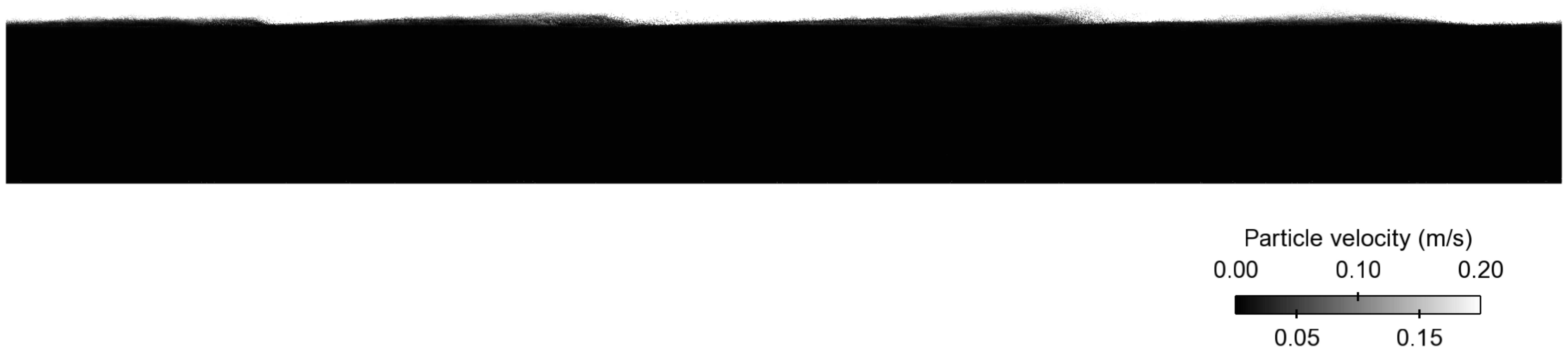}}
}
\end{minipage}
\caption{Top and side views of the instantaneous particle configuration for case 1, colored by the particle velocity.}
\label{config_case1}
\end{figure}

\begin{figure}
\begin{minipage}[t!]{70mm}
\centering{
\subfigure{\includegraphics[width=120mm]{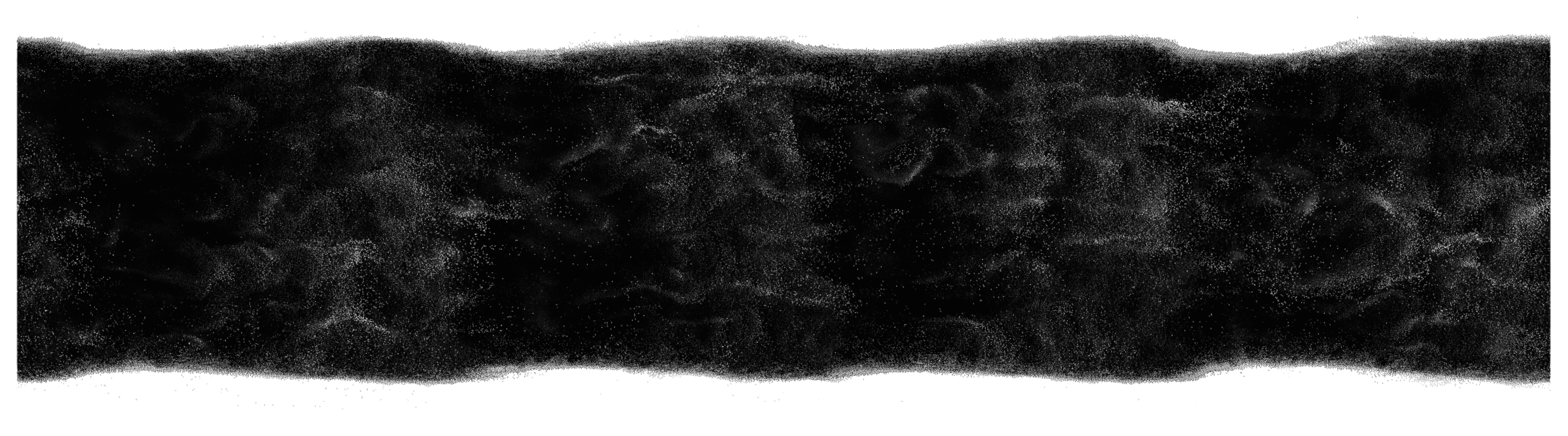}}
\subfigure{\includegraphics[width=120mm]{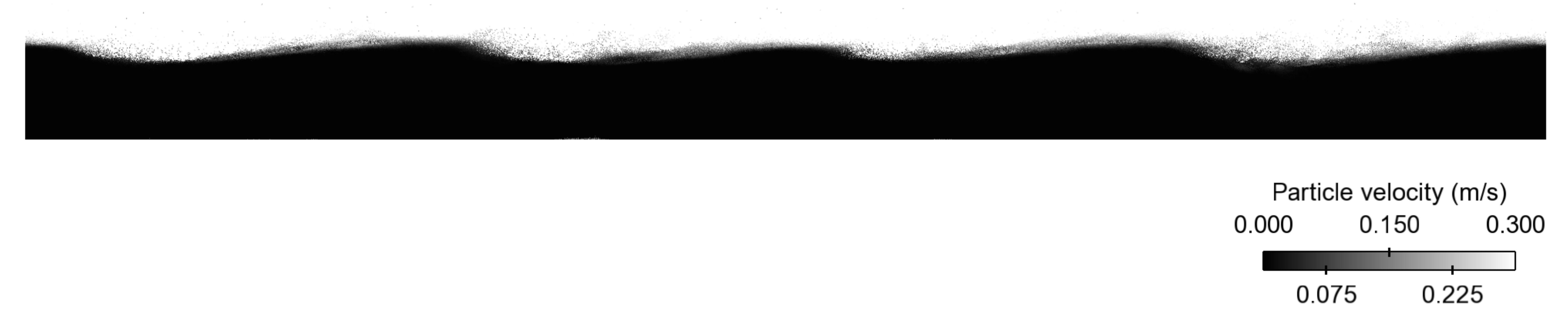}}
}
\end{minipage}
\caption{Top and side views of the instantaneous particle configuration for case 2, colored by the particle velocity.}
\label{config_case2}
\end{figure}

\begin{figure}
\begin{minipage}[t!]{70mm}
\centering{
\subfigure{\includegraphics[width=120mm]{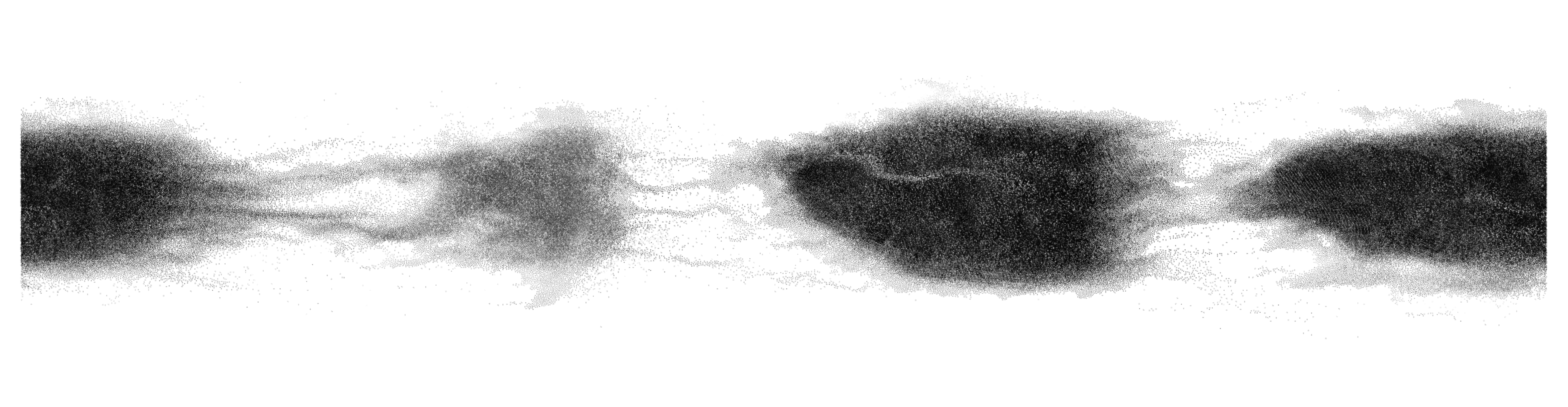}}
\subfigure{\includegraphics[width=120mm]{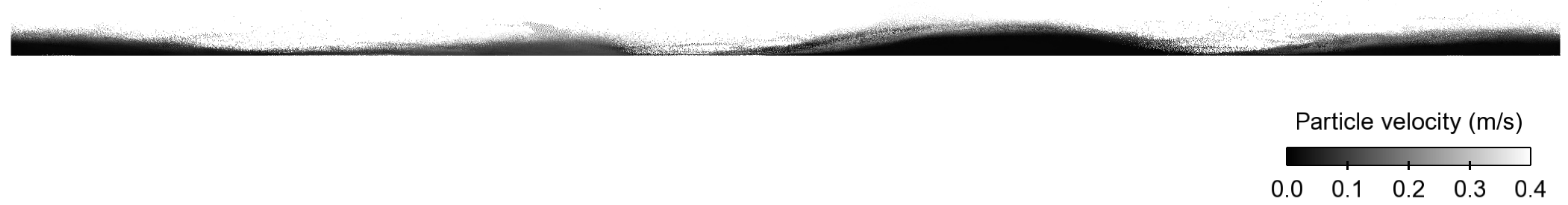}}
}
\end{minipage}
\caption{Top and side views of the instantaneous particle configuration for case 3, colored by the particle velocity.}
\label{config_case3}
\end{figure}

Instantaneous snapshots of the particle configuration are plotted in figures \ref{config_case1} to \ref{config_case3}. At low liquid flow rates, only small amplitude dunes are formed as evident from the side view in figure \ref{config_case1}. As the flow rate is increased, the ridges and troughs are more clearly visible in the side view for both case 2 and 3 as shown in figures \ref{config_case2} and \ref{config_case3}. For case 3 with the highest liquid flow rate, the patterns show that there are regions at the bottom of the pipe where there are no particles (in the troughs). This is consistent with the experimental observations of \cite{ouriemi:2009b}. 

\begin{figure}
\begin{minipage}[t!]{100mm}
\centering{
\subfigure[Case 1]{\includegraphics[width=100mm]{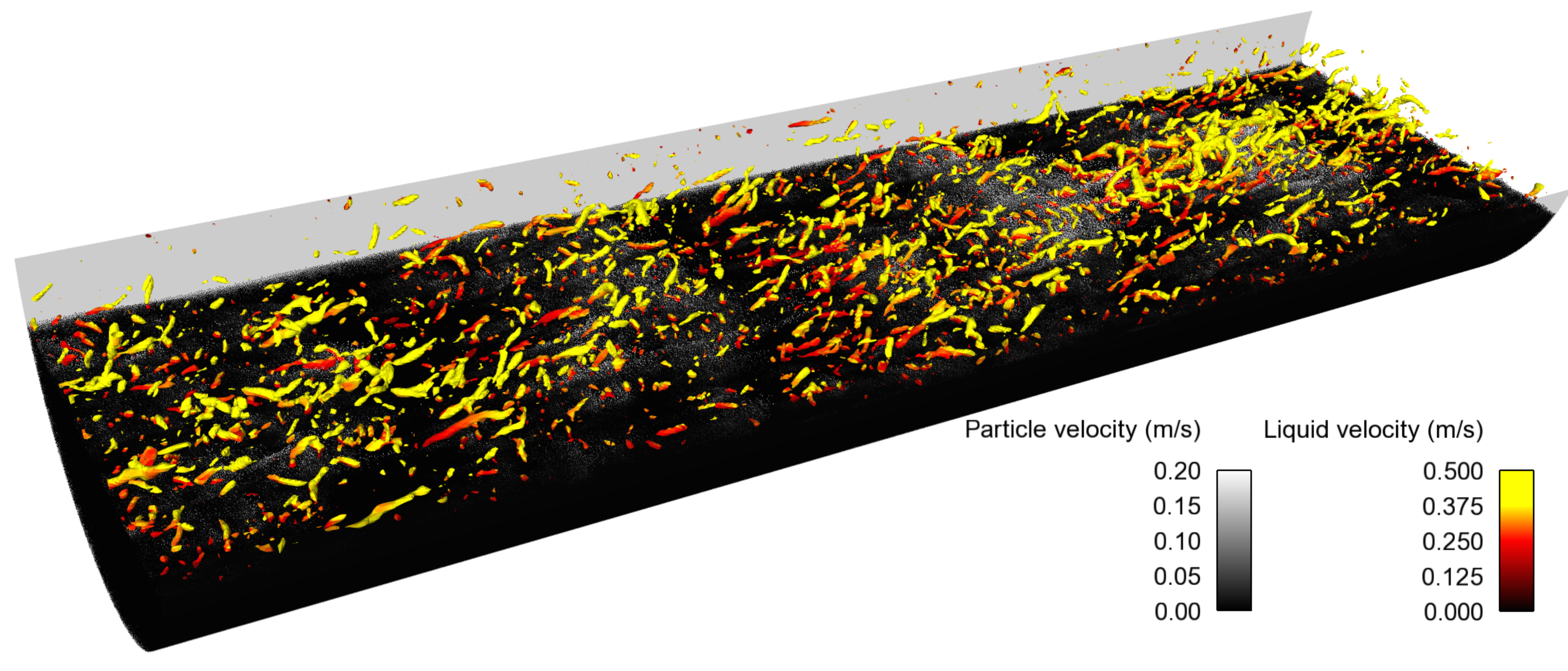}}
}
\end{minipage}
\hfill
\begin{minipage}[t!]{100mm}
\centering{
\subfigure[Case 2]{\includegraphics[width=100mm]{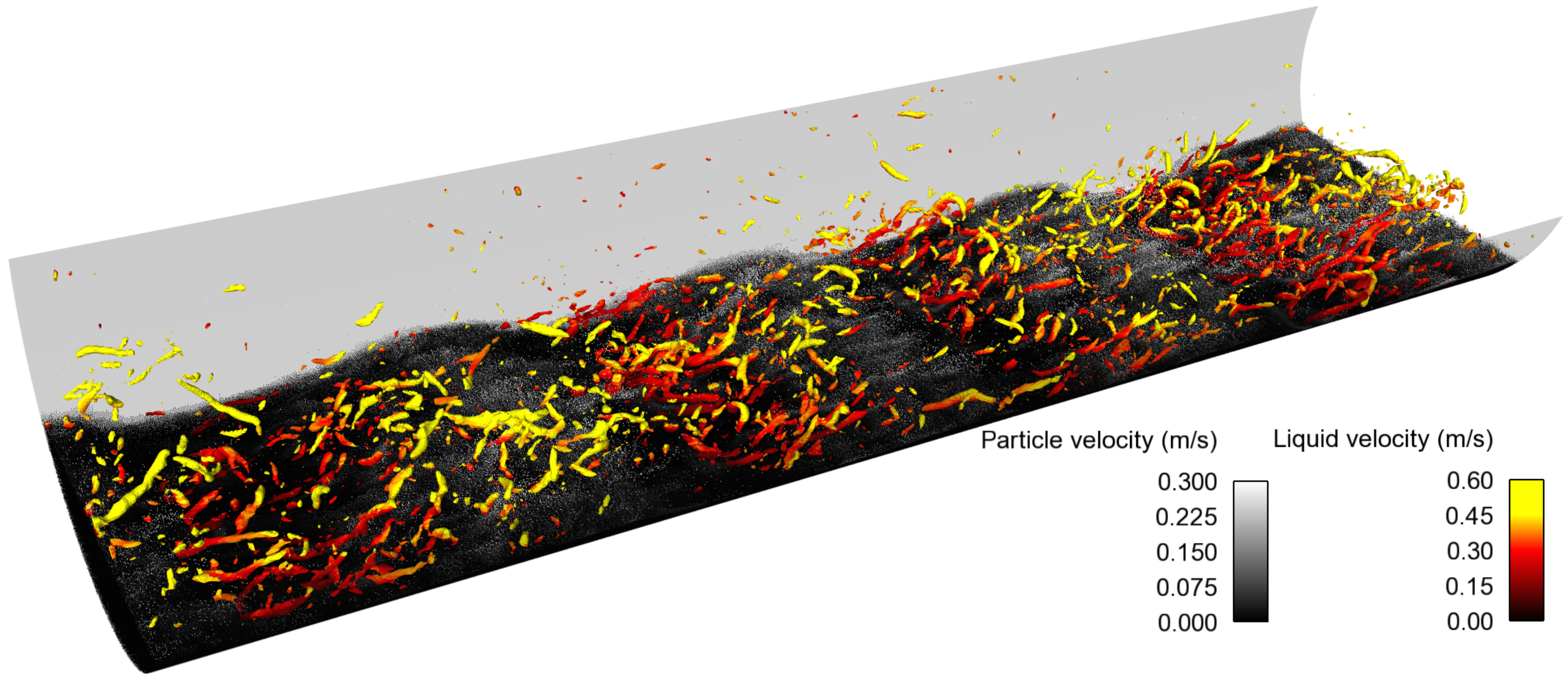}}
}
\end{minipage}
\hfill
\begin{minipage}[t!]{100mm}
\centering{
\subfigure[Case 3]{\includegraphics[width=100mm]{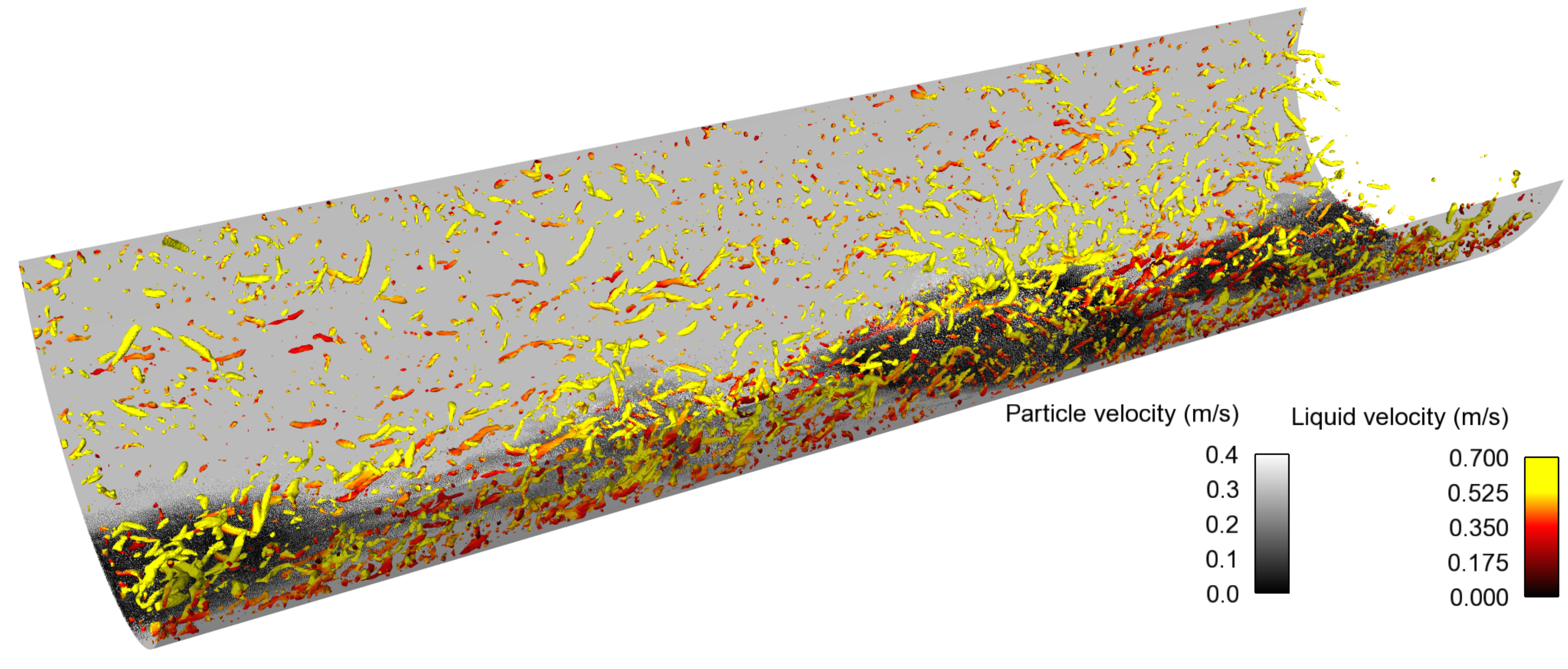}}
}
\end{minipage}
\caption{Vortical structures detected using iso-surfaces of $Q$-criterion}
\label{vorticalStructures}
\end{figure}

To further understand and characterize the dunes, vortical structures identified through iso-surfaces of $Q$-criterion are plotted in figure \ref{vorticalStructures}. $Q$ is the second invariant of the velocity gradient tensor, defined as
\begin{equation}
Q=(|\bm{\Omega}|^{2}-|\bm{S}|^{2})
\end{equation}
where $\bm{\Omega}$ is the rate of rotation tensor and $\bm{S}$ is the rate of strain tensor. Positive $Q$ isosurfaces can therefore be used to identify coherent vortices, as regions where the rate of rotation is greater than the rate of strain \citep{hunt:1988}.  In case 1, there are only small amplitude dunes and hence the vortical structures do not seem to be noticeably influenced by these patterns. In case 2, however, vortical structures are coupled to the dune patterns. These coherent structures essentially indicate flow separation in the troughs. Following classification proposed by \cite{ouriemi:2009b}, the dune patterns observed in case 2 can be denoted as ``vortex dunes''. Case 3 also shows vortical structures, but the particles are present very close to the bottom wall of the pipe, it is difficult to visualize the structures above the dune patterns.

Overall, the formation of small dunes for case 1, and the formation of vortex dunes for the other two cases suggest that the present simulations reproduce appropriately the expected behavior from the phase diagrams.

\subsection{Space-time evolution of the liquid-bed interface}
\begin{figure}
\begin{minipage}[t!]{44mm}
\centering{
\subfigure[Case 1]{\includegraphics[width=44mm]{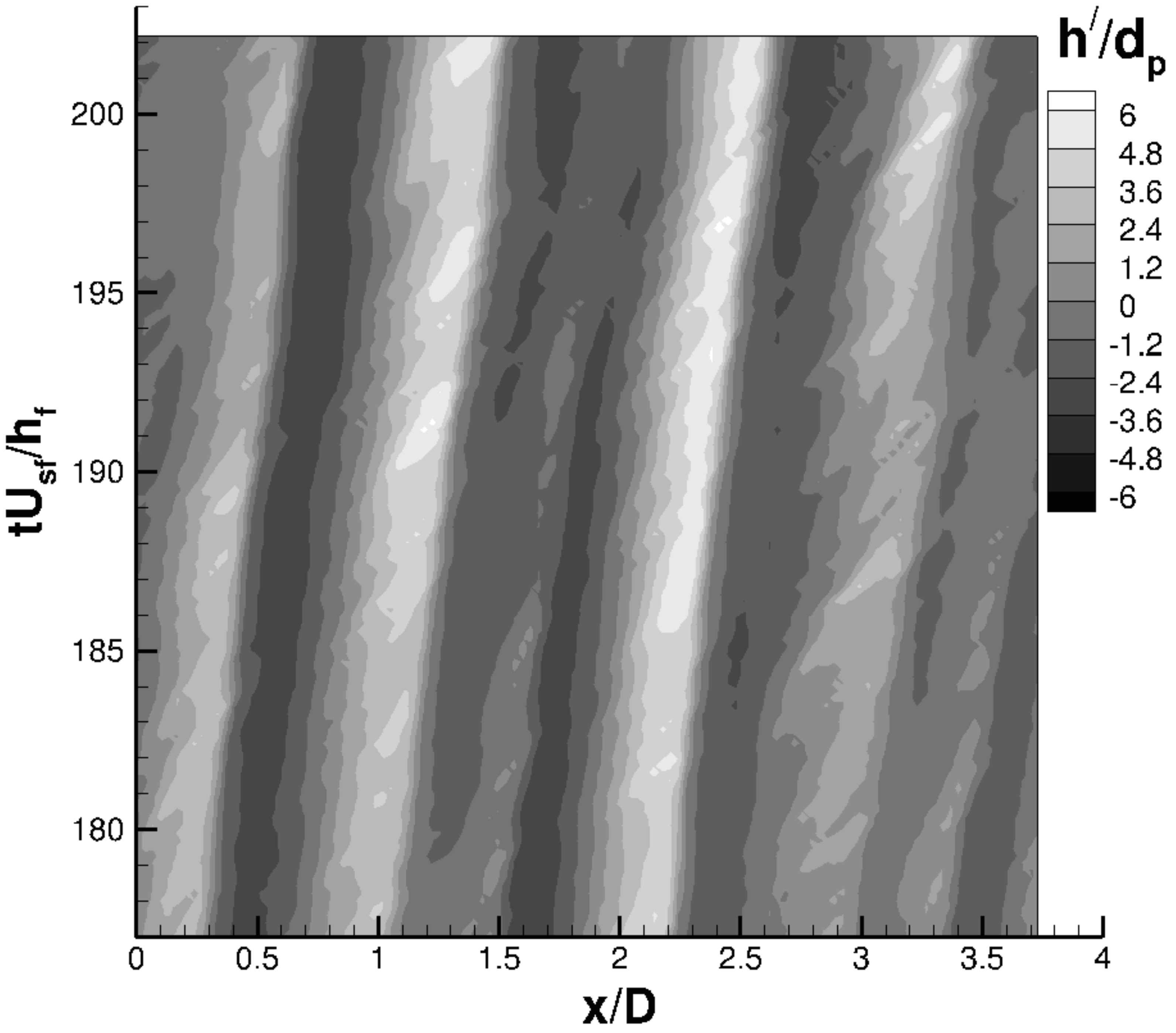}}
}
\end{minipage}
\hfill
\begin{minipage}[t!]{44mm}
\centering{
\subfigure[Case 2]{\includegraphics[width=44mm]{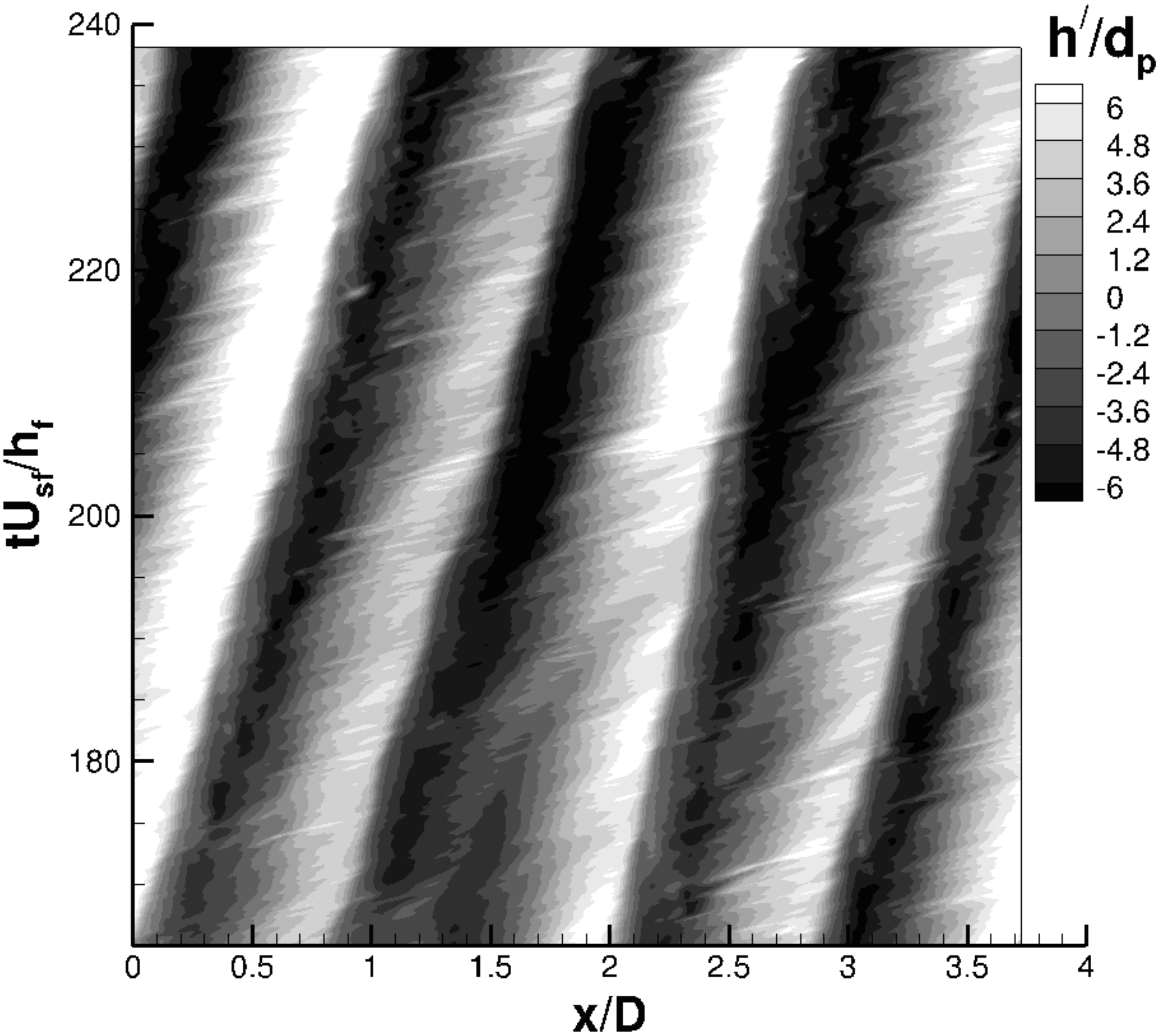}}
}
\end{minipage}
\hfill
\begin{minipage}[t!]{44mm}
\centering{
\subfigure[Case 3]{\includegraphics[width=44mm]{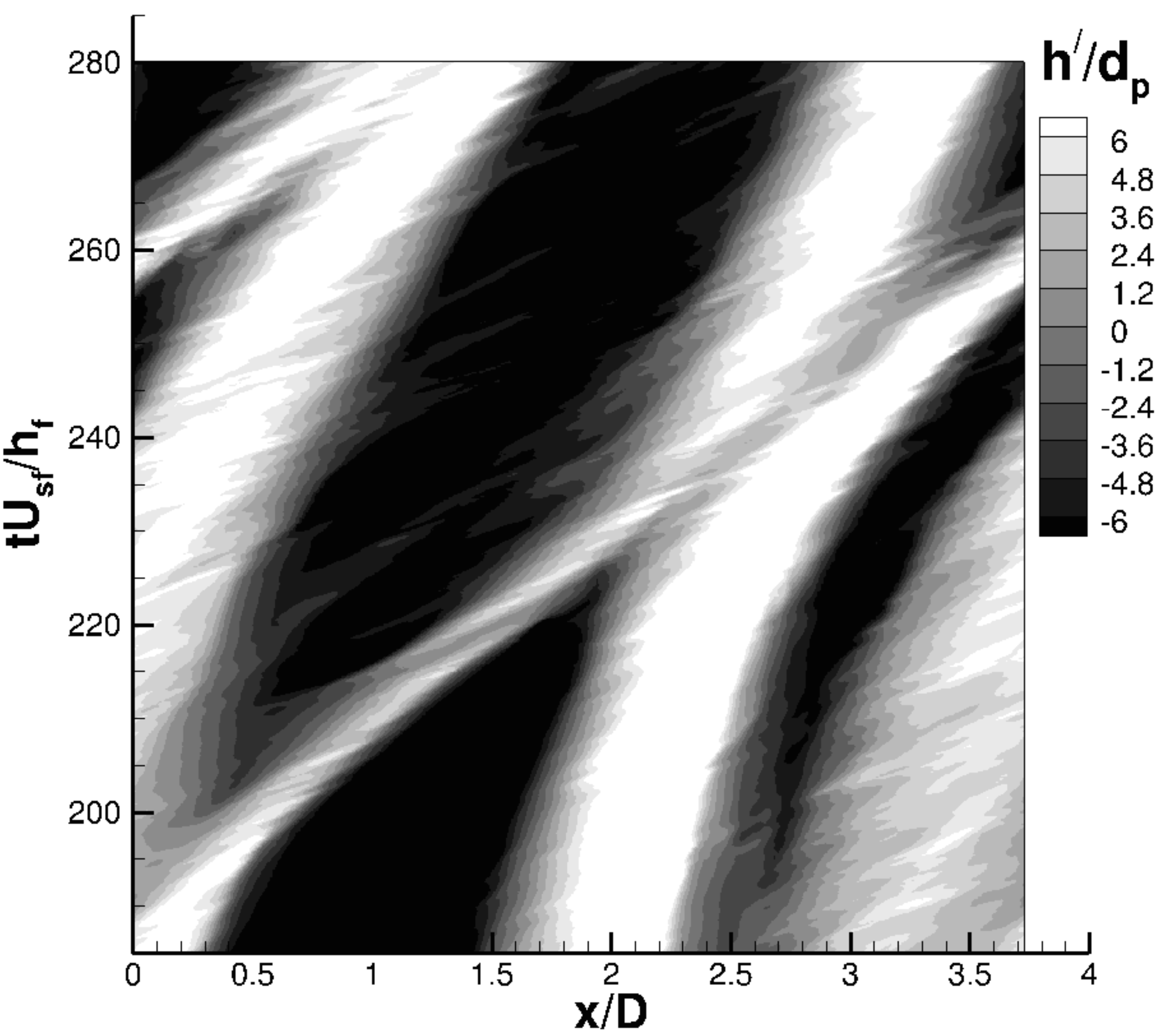}}
}
\end{minipage}
\caption{Space-time diagram showing evolution of the bed patterns.}
\label{Space-time diagram}
\end{figure}

To further elucidate the space-time evolution of the patterns of liquid-solid interface, the fluctuation in the height of liquid-bed interface is calculated using
\begin{equation}
h^{\prime}(x,t) = h_{i}(x,t) - \langle h_{i} \rangle_x (t)
\end{equation}
where $\la \cdot \ra_{x}$ denotes averaging in the streamwise direction. Space-time plots of $h^{\prime}$ normalized by $d_p$ are shown in figure \ref{Space-time diagram}. The ridges and troughs that are observed in this figure again indicate dune formation. In the lowest liquid flow rate case, the fluctuations in the height of the liquid-bed interface are weak. For the other two cases, the fluctuations have a larger amplitude. The convection speed of the patterns can be calculated by looking at the slope of the line determined by the locus of the maximum of the height of the liquid-bed interface. In the final period of the simulation, that convective speed in both case 1 and case 2 is approximately 3\% of $U_{sf}$. In case 3, the patterns exhibit interesting dynamics such as merging and branching, and the convective dune speed in the final period for this case is approximately 5.5\% of $U_{sf}$. The fact that the convection speed is highest in case 3 might be due to the fact that in that case, the dunes are sliding and rolling at the bottom of the pipe.

\section{Concluding remarks}
The dynamics of liquid-solid slurry flows through a horizontal pipe has been investigated using a highly resolved Euler-Lagrange LES strategy. Three liquid flow rates were considered, leading to different slurry dynamics that are consistent with the existing literature. The incipient erosion of the particle bed, referred to as region I, is determined by the bed shear stress. Once the particles are eroded, they move in a thin, dense layer referred to as region II where particle-particle collisions play a significant role. Particles can then be fed to the vortical structures of the turbulent flow, leading to few fast-moving particles in region III. The dense particle dynamics, interacting with the liquid phase turbulence, leads to the formation of various patterns above the static bed. The evolution of these patterns in space and time is presented and the propagation speed is calculated. The mean streamwise pressure gradient from numerical simulations agree with the experiments within engineering accuracy levels. The vertical profiles for the mean particle velocity and concentration are extracted from the simulations and discussed. To our knowledge, this is the first work demonstrating the capability of accurately predicting the critical deposition velocity. These simulations show that Euler-Lagrange LES can be used as a viable tool in gaining physical insight of such complex liquid-solid flows.

\bibliographystyle{elsarticle-harv}

\bibliography{SlurryPaper}

\begin{thebibliography}{36}
\expandafter\ifx\csname natexlab\endcsname\relax\def\natexlab#1{#1}\fi
\expandafter\ifx\csname url\endcsname\relax
  \def\url#1{\texttt{#1}}\fi
\expandafter\ifx\csname urlprefix\endcsname\relax\def\urlprefix{URL }\fi

\bibitem[{Bagnold(1966)}]{bagnold:1966}
Bagnold, R.~A., 1966. An approach to the sediment transport problem from
  general physics. Geological Survey Professional Paper 422-I.

\bibitem[{Buffington and Montgomery(1997)}]{buffington:1997}
Buffington, J.~M., Montgomery, D.~R., 1997. A systematic analysis of eight
  decades of incipient motion studies, with special reference to gravel-bedded
  rivers. Water Resources Research 33~(8), 1993--2029.

\bibitem[{Capecelatro and Desjardins(2013{\natexlab{a}})}]{capecelatro:2013a}
Capecelatro, J., Desjardins, O., 2013{\natexlab{a}}. An {E}uler--{L}agrange
  strategy for simulating particle-laden flows. Journal of Computational
  Physics 238, 1 -- 31.

\bibitem[{Capecelatro and Desjardins(2013{\natexlab{b}})}]{capecelatro:2013b}
Capecelatro, J., Desjardins, O., 2013{\natexlab{b}}. {E}ulerian--{L}agrangian
  modeling of turbulent liquid--solid slurries in horizontal pipes.
  International Journal of Multiphase Flow 55, 64 -- 79.

\bibitem[{Charru(2006)}]{charru:2006}
Charru, F., 2006. Selection of the ripple length on a granular bed sheared by a
  liquid flow. Physics of Fluids 18~(121508), 1--9.

\bibitem[{Church(2006)}]{church:2006}
Church, M., 2006. Bed material transport and the morphology of alluvial river
  channels. Annual Review of Earth and Planetary Sciences 34~(1), 325--354.

\bibitem[{Coleman and Nikora(2009)}]{coleman:2009}
Coleman, S., Nikora, V.~I., 2009. Bed and flow dynamics leading to
  sediment-wave initiation. Water Resources Research 45~(4), 1--12.

\bibitem[{Cundall and Strack(1979)}]{cundall:1979}
Cundall, P., Strack, O., 1979. A discrete numerical model for granular
  assemblies. Geotechnique 29~(1), 47--65.

\bibitem[{Dahl et~al.(2003)Dahl, Ladam, Unander, and Onsrud}]{dahl:2003}
Dahl, A.~M., Ladam, Y., Unander, T., Onsrud, G., 2003. {SINTEF-IFE} {S}and
  transport 2001-2003. In: SINTEF Internal Report.

\bibitem[{Dancey et~al.(2002)Dancey, Diplas, Papanicolaou, and
  Bala}]{dancey:2002}
Dancey, C.~L., Diplas, P., Papanicolaou, A., Bala, M., 2002. Probability of
  individual grain movement and threshold condition. Journal of Hydraulic
  Engineering 128~(12), 1069--1075.

\bibitem[{Danielson(2007)}]{danielson:2007}
Danielson, T.~J., 30 April - 3 May 2007. Sand transport modeling in multiphase
  pipelines. Houston, Texas, U.S.A.

\bibitem[{Desjardins et~al.(2008)Desjardins, Blanquart, Balarac, and
  Pitsch}]{desjardins:2008}
Desjardins, O., Blanquart, G., Balarac, G., Pitsch, H., 2008. {High order
  conservative finite difference scheme for variable density low {M}ach number
  turbulent flows}. Journal of Computational Physics 227~(15), 7125--7159.

\bibitem[{Dur{\'a}n et~al.(2012)Dur{\'a}n, Andreotti, and Claudin}]{duran:2012}
Dur{\'a}n, O., Andreotti, B., Claudin, P., 2012. Numerical simulation of
  turbulent sediment transport, from bed load to saltation. Physics of Fluids
  (1994-present) 24~(10), 103306.

\bibitem[{Germano et~al.(1991)Germano, Piomelli, Moin, and
  Cabot}]{germano:1991}
Germano, M., Piomelli, U., Moin, P., Cabot, W.~H., 1991. A dynamic
  subgrid-scale eddy viscosity model. Physics of Fluids A: Fluid Dynamics 3,
  1760.

\bibitem[{Gibilaro et~al.(2007)Gibilaro, Gallucci, Di~Felice, and
  Pagliai}]{gibilaro:2007}
Gibilaro, L., Gallucci, K., Di~Felice, R., Pagliai, P., 2007. On the apparent
  viscosity of a fluidized bed. Chemical engineering science 62~(1-2),
  294--300.

\bibitem[{Hunt et~al.(1988)Hunt, Wray, and Moin}]{hunt:1988}
Hunt, J. C.~R., Wray, A., Moin, P., 1988. Eddies, stream, and convergence zones
  in turbulent flows. Center for Turbulence Research Report CTR-S88.

\bibitem[{Jenkins and Hanes(1998)}]{jenkins:1998}
Jenkins, J.~T., Hanes, D.~M., 9 1998. Collisional sheet flows of sediment
  driven by a turbulent fluid. Journal of Fluid Mechanics 370, 29--52.

\bibitem[{Kidanemariam and Uhlmann(2014)}]{kidanemariam:2014}
Kidanemariam, A.~G., Uhlmann, M., 2014. Direct numerical simulation of pattern
  formation in subaqueous sediment. Journal of Fluid Mechanics 750~(R2), 1--13.

\bibitem[{Langlois and Valance(2007)}]{langlois:2007}
Langlois, V., Valance, A., 2007. Initiation and evolution of current ripples on
  a flat sand bed under turbulent water flow. The European Physical Journal E:
  Soft Matter and Biological Physics 22~(3), 201--208.

\bibitem[{Lilly(1992)}]{lilly:1992}
Lilly, D., 1992. A proposed modification of the {G}ermano subgrid-scale closure
  method. Physics of Fluids A: Fluid Dynamics 4, 633.

\bibitem[{McCaslin and Desjardins(2014)}]{mccaslin:2014}
McCaslin, J.~O., Desjardins, O., 2014. Numerical investigation of gravitational
  effects in horizontal annular liquid-gas flow. International Journal of
  Multiphase Flow 67, 88--105.

\bibitem[{Meneveau et~al.(1996)Meneveau, Lund, and Cabot}]{meneveau:1996}
Meneveau, C., Lund, T., Cabot, W., 1996. A {L}agrangian dynamic subgrid-scale
  model of turbulence. Journal of Fluid Mechanics 319~(1), 353--385.

\bibitem[{Nielsen(1992)}]{nielsen:1992}
Nielsen, P., 1992. Coastal bottom boundary layers and sediment transport.
  Advanced Series on Ocean Engineering: Volume 4. World Scientific.

\bibitem[{Ouriemi et~al.(2009{\natexlab{a}})Ouriemi, Aussillous, and
  Guazzelli}]{ouriemi:2009a}
Ouriemi, M., Aussillous, P., Guazzelli, E., 2009{\natexlab{a}}. Sediment
  dynamics. {P}art 1. {B}ed-load transport by laminar shearing flows. Journal
  of Fluid Mechanics 636, 295--319.

\bibitem[{Ouriemi et~al.(2009{\natexlab{b}})Ouriemi, Aussillous, and
  Guazzelli}]{ouriemi:2009b}
Ouriemi, M., Aussillous, P., Guazzelli, E., 2009{\natexlab{b}}. Sediment
  dynamics. {P}art 2. {D}une formation in pipe flow. Journal of Fluid Mechanics
  636, 321--336.

\bibitem[{Ouriemi et~al.(2007)Ouriemi, Aussillous, Medale, Peysson, and
  Guazzelli}]{ouriemi:2007}
Ouriemi, M., Aussillous, P., Medale, M., Peysson, Y., Guazzelli, {\'E}., 2007.
  Determination of the critical {S}hields number for particle erosion in
  laminar flow. Physics of Fluids 19~(6), 61706--63100.

\bibitem[{Paintal(1971)}]{paintal:1971}
Paintal, A., 1971. Concept of critical shear stress in loose boundary open
  channels. Journal of Hydraulic Research 9~(1), 91--113.

\bibitem[{Papnicolaou et~al.(2002)Papnicolaou, Diplas, Evaggelopoulos, and
  Fotopoulos}]{papanicolaou:2002}
Papnicolaou, A.~N., Diplas, P., Evaggelopoulos, N., Fotopoulos, S., 2002.
  Stochastic incipient motion criterion for spheres under various bed packing
  conditions. Journal of Hydraulic Engineering 128~(4), 369--380.

\bibitem[{Peysson et~al.(2009)Peysson, Ouriemi, Medale, Aussillous, and
  Guazzelli}]{peysson:2009}
Peysson, Y., Ouriemi, M., Medale, M., Aussillous, P., Guazzelli, {\'E}., 2009.
  Threshold for sediment erosion in pipe flow. International Journal of
  Multiphase Flow 35~(6), 597--600.

\bibitem[{Raudkivi(1997)}]{raudkivi:1997}
Raudkivi, A., 1997. Ripples on stream bed. Journal of Hydraulic Engineering
  123~(1), 58--64.

\bibitem[{Tenneti et~al.(2011)Tenneti, Garg, and Subramaniam}]{tenneti:2011}
Tenneti, S., Garg, R., Subramaniam, S., 2011. Drag law for monodisperse
  gas-solid systems using particle-resolved direct numerical simulation of flow
  past fixed assemblies of spheres. International Journal of Multiphase Flow
  37~(9), 1072--1092.

\bibitem[{Van~Rijn(1984)}]{vanrijn:1984}
Van~Rijn, L.~C., 1984. Sediment transport, {P}art {II}: {S}uspended load
  transport. Journal of Hydraulic Engineering 110~(11), 1613--1641.

\bibitem[{Vanoni(1946)}]{vanoni:1946}
Vanoni, V.~A., 1946. Transportation of suspended sediment by water.
  Transactions of the American Society of Civil Engineers 111, 67--113.

\bibitem[{Wilson(1989)}]{wilson:1989}
Wilson, K., 1989. Friction of wave-induced sheet flow. Coastal Engineering 12,
  371--379.

\bibitem[{Yalin(1977)}]{yalin:1977}
Yalin, M.~S., 1977. Mechanics of sediment transport. Pergamon Press.

\bibitem[{Yang et~al.(2006)Yang, Ladam, Laux, Danielson, and
  Leporcher}]{yang:2006}
Yang, Z.~L., Ladam, Y., Laux, H., Danielson, T.~J., Leporcher, E., 2006.
  Dynamic simulation of sand transport in a pipeline. In: 5th North American
  Conference on Multiphase Technology. Banff, Canada.

\end{thebibliography}

\end{document}